\documentclass{aa}
\usepackage{graphicx}
\usepackage{lscape}
\usepackage{subfigure}
\usepackage{natbib}
\usepackage[varg]{txfonts}
\usepackage{longtable}
\usepackage{booktabs}
\usepackage[normalem]{ulem} 
\usepackage{epstopdf}
\usepackage{xcolor}
\usepackage{soul}
\usepackage{makecell}
\usepackage{multicol,tabularx,capt-of}
\usepackage{placeins}
\usepackage{scalerel}
\usepackage{tikz}

\usetikzlibrary{svg.path}

\defcitealias{Sanchez-Saez21a}{SS21a}

\definecolor{orcidlogocol}{HTML}{A6CE39}
\tikzset{
  orcidlogo/.pic={
    \fill[orcidlogocol] svg{M256,128c0,70.7-57.3,128-128,128C57.3,256,0,198.7,0,128C0,57.3,57.3,0,128,0C198.7,0,256,57.3,256,128z};
    \fill[white] svg{M86.3,186.2H70.9V79.1h15.4v48.4V186.2z}
                 svg{M108.9,79.1h41.6c39.6,0,57,28.3,57,53.6c0,27.5-21.5,53.6-56.8,53.6h-41.8V79.1z M124.3,172.4h24.5c34.9,0,42.9-26.5,42.9-39.7c0-21.5-13.7-39.7-43.7-39.7h-23.7V172.4z}
                 svg{M88.7,56.8c0,5.5-4.5,10.1-10.1,10.1c-5.6,0-10.1-4.6-10.1-10.1c0-5.6,4.5-10.1,10.1-10.1C84.2,46.7,88.7,51.3,88.7,56.8z};
  }
}

\newcommand\orcidicon[1]{\href{https://orcid.org/#1}{\mbox{\scalerel*{
\begin{tikzpicture}[yscale=-1,transform shape]
\pic{orcidlogo};
\end{tikzpicture}
}{|}}}}

\usepackage{hyperref} 
\hypersetup{
    colorlinks=true,
    citecolor=blue,
    linkcolor=blue,
    urlcolor=blue,
    }
\makeatletter
\renewcommand*\aa@pageof{, page \thepage{} of \pageref*{LastPage}}
\makeatother

\bibpunct{(}{)}{;}{a}{}{,} 
\usepackage{amstext} 
\definecolor{cadmiumred}{rgb}{0.89, 0.0, 0.13}

\definecolor{ste}{rgb}{0., 0.26, 0.15}

\begin{document} 

   \title{SDSS1335+0728: The awakening  of a $\sim 10^6 M_{\odot}$ black hole\thanks{Based on observations collected at the European Southern Observatory under ESO programme 109.24F5.001. The SOAR/Goodman and Keck/LRIS spectra are available in electronic form at the CDS via anonymous ftp to cdsarc.u-strasbg.fr (130.79.128.5) or via http://cdsweb.u-strasbg.fr/cgi-bin/qcat?J/A+A/}}

  \author{
   P. S\'anchez-S\'aez\inst{1,2\orcidicon{0000-0003-0820-4692}},
   L. Hern\'andez-Garc\'ia\inst{2,3\orcidicon{0000-0002-8606-6961}},
   S. Bernal\inst{3,4},
   A. Bayo\inst{1\orcidicon{0000-0001-7868-7031}},
   G. Calistro Rivera\inst{1\orcidicon{0000-0003-0085-6346}},
   F. E. Bauer\inst{5,6,2,7\orcidicon{0000-0002-8686-8737}},
   C. Ricci\inst{8,9\orcidicon{0000-0001-5231-2645}},
   A. Merloni\inst{10},
   M. J. Graham\inst{11\orcidicon{0000-0002-3168-0139}},
   R. Cartier\inst{12,8},
   P. Ar\'evalo\inst{3,4\orcidicon{0000-0001-8503-9809}},
   R. J. Assef\inst{8\orcidicon{0000-0002-9508-3667}},
   A. Concas\inst{1,13},
   D. Homan\inst{14},
   M. Krumpe\inst{14},
   P. Lira\inst{15,4\orcidicon{0000-0003-1523-9164}},
   A. Malyali\inst{10},
   M.L. Martínez-Aldama\inst{16},
   A. M. Mu\~noz Arancibia\inst{2,17\orcidicon{0000-0002-8722-516X}},
   A. Rau\inst{10\orcidicon{0000-0001-5990-6243}},
   G. Bruni\inst{18\orcidicon{0000-0002-5182-6289}},
   F. F\"orster\inst{19,2,17,15\orcidicon{0000-0003-3459-2270}},
   M. Pavez-Herrera\inst{2},
   D. Tub\'in-Arenas\inst{14\orcidicon{0000-0002-2688-7960}}
   M. Brightman\inst{20\orcidicon{0000-0002-8147-2602}}
   }

   \titlerunning{SDSS1335+0728: The awakening of a $\sim 10^6 M_{\odot}$ black hole}
   \authorrunning{P. S\'anchez-S\'aez et al.}

\institute{
European Southern Observatory, Karl-Schwarzschild-Strasse 2, 85748 Garching bei München, Germany; \href{mailto:psanchez@eso.org}{psanchez@eso.org}
\and
Millennium Institute of Astrophysics (MAS), Nuncio Monse\~nor Sotero Sanz 100, Of. 104, Providencia, Santiago, Chile
\and
Instituto de F\'isica y Astronom\'ia, Facultad de Ciencias,Universidad de Valpara\'iso, Gran Breta\~na No. 1111, Playa Ancha, Valpara\'iso, Chile
\and
Millennium Nucleus on Transversal Research and Technology to Explore Supermassive Black Holes (TITANS), 4030000 Concepción, Chile
\and
Instituto de Astrof{\'{\i}}sica, Facultad de F{\'{i}}sica, Pontificia Universidad Cat{\'{o}}lica de Chile, Casilla 306, Santiago 22, Chile
\and
Centro de Astroingenier{\'{\i}}a, Pontificia Universidad Cat{\'{o}}lica de Chile, Av. Vicu\~{n}a Mackenna 4860, 7820436 Macul, Santiago, Chile
\and
Space Science Institute, 4750 Walnut Street, Suite 205, Boulder, Colorado 80301
\and
Instituto de Estudios Astrofísicos, Facultad de Ingeniería y Ciencias, Universidad Diego Portales, Av. Ejército Libertador 441, Santiago, Chile
\and
Kavli Institute for Astronomy and Astrophysics, Peking University, Beĳing 100871, China
\and
Max-Planck-Institut f\"ur Extraterrestrische Physik, Gie{\ss}enbachstra{\ss}e, D-85748 Garching, Germany
\and
California Institute of Technology, 1200 E. California Blvd, Pasadena, CA 91125, USA
\and
Gemini Observatory, NSF’s National Optical-Infrared Astronomy Research Laboratory, Casilla 603, La Serena, Chile
\and
INAF - Osservatorio Astrofisico di Arcetri, Largo E. Fermi 5, 50125, Firenze, Italy
\and
Leibniz-Institut für Astrophysik Potsdam, An der Sternwarte 16, 14482 Potsdam, Germany
\and
Departamento de Astronom\'ia, Universidad de Chile, Casilla 36D, Santiago, Chile
\and
Astronomy Department, Universidad de Concepción, Barrio Universitario S/N, Concepción 4030000, Chile
\and
Center for Mathematical Modeling, University of Chile, AFB170001, Chile
\and
INAF -- Institute for Space Astrophysics and Planetology, via del Fosso del Cavaliere, 100, I-00133 Roma, Italy
\and
Data and Artificial Intelligence Initiative (ID\&IA), University of Chile, Santiago, Chile
\and
Cahill Center for Astrophysics, California Institute of Technology, 1216 East California Boulevard, Pasadena, CA 91125, USA
}

   \date{}
  \abstract
   {The early-type galaxy SDSS J133519.91+072807.4 (hereafter SDSS1335+0728), which had exhibited no prior optical variations during the preceding two decades, began showing significant nuclear variability in the \textit{Zwicky} Transient Facility (ZTF) alert stream from December 2019 (as ZTF19acnskyy). This variability behaviour, coupled with the host-galaxy properties, suggests that SDSS1335+0728 hosts a $\sim 10^6 M_{\odot}$ black hole (BH) that is currently in the process of `turning on'.}
   { We present a multi-wavelength photometric analysis and spectroscopic follow-up performed with the aim of better understanding the origin of the nuclear variations detected in SDSS1335+0728. }
   {We used archival photometry (from WISE, 2MASS, SDSS, GALEX, eROSITA) and spectroscopic data (from SDSS and LAMOST) to study the state of SDSS1335+0728 prior to December 2019, and new observations from \emph{Swift}, SOAR/Goodman, VLT/X-shooter, and Keck/LRIS taken after its turn-on to characterise its current state. We analysed the variability of SDSS1335+0728 in the X-ray/UV/optical/mid-infrared range, modelled its spectral energy distribution prior to and after December 2019, and studied the evolution of its UV/optical spectra. } 
  {From our multi-wavelength photometric analysis, we find that: (a) {since 2021, the UV flux (from \emph{Swift}/UVOT observations) is four times brighter than the flux reported by GALEX in 2004}; (b) since June 2022, the mid-infrared flux has risen more than two times, and the W1$-$W2 WISE colour has become redder; and (c) since February 2024, the source has begun showing X-ray emission. From our spectroscopic follow-up, we see that (i) the narrow emission line ratios are now consistent with a more energetic ionising continuum; (ii) broad emission lines are not detected; and (iii) the [OIII] line increased its flux $\sim 3.6$ years after the first ZTF alert, which implies a relatively compact narrow-line-emitting region. }
   {We conclude that the variations observed in SDSS1335+0728 could be either explained by a $\sim 10^6 M_{\odot}$ AGN that is just turning on or by an exotic tidal disruption event (TDE). If the former is true, SDSS1335+0728 is one of the strongest cases of an AGN observed in the process of activating. If the latter were found to be the case, it would correspond to the longest and faintest TDE ever observed (or another class of still unknown nuclear transient). Future observations of SDSS1335+0728  are crucial to further understand its behaviour.}  


\maketitle

\section{Introduction}\label{section:intro}
All-sky time-domain surveys, such as the \textit{Zwicky} Transient Facility (ZTF; \citealt{Bellm19}) and the upcoming \textit{Vera C. Rubin} Observatory Legacy Survey of Space and Time (LSST; \citealt{LSST}), which are generating alert streams in real time, are facilitating the identification of nuclear transients in galaxies hosting supermassive black holes (SMBHs). These events include tidal disruption events (TDEs); the disruption of a star 
 when it approximates a  black hole (BH) with a mass of $\lesssim 10^8  M_{\odot}$ \citep{Rees88,vanVelzen20}; active galactic nuclei (AGN) with changing-state events (CSAGNs, also known as changing-look AGNs); sources that drastically change their accretion state, which could lead to changes in their classification as type 1 or type 2 AGN (with or without broad emission lines) within a timescale of months or years \citep{Ricci23}; AGNs with anomalous flaring activity (e.g. \citealt{Trakhtenbrot19NatAs,Frederick21}); and the recently discovered ambiguous nuclear transients (ANTs; e.g. \citealt{Hinkle22} and references therein).

Nuclear transients could also correspond to AGN ignition events in formerly quiescent galaxies (e.g. \citealt{Arevalo24}). The duty cycle, meaning the time an AGN is active, is estimated to range between 10$^4$ and 10$^7$ years (for classical AGNs), and during this period, while a SMBH is switching on and off (i.e., going through episodes of lower and higher activity), the expected brightness change can be of several orders of magnitude \citep{Hickox14}. However, from the Soltan argument \citep{Soltan82}, we know most of the mass of the SMBHs we observe has already been obtained through accretion, and hence we do not expect to see many AGNs today in the process of turning on. However, intermediate-mass BH (IMBHs) with masses of $10^{4} - 10^5  M_{\odot}$ or low-mass SMBHs with masses of $10^5 \sim 10^6  M_{\odot}$ \citep{Greene20} may still have masses close to those of their seeds (with masses of $M \thickapprox 10^{2} \: M_{\odot}$ for light seeds, and masses of $M \thickapprox 10^{3} - 10^5 M_{\odot}$ for heavy seeds) and, if so, they must not have gone through many duty cycles and their AGN activity may be incipient. Accordingly, we would expect to find IMBHs and low-mass SMBHs today that are just starting to become active, and that will eventually grow to become SMBHs.

Deciphering whether a nuclear transient corresponds to a TDE, a CSAGN, a flaring AGN, or a turning-on AGN event could be challenging, as the number of known sources presenting these behaviours is still low, and the timescales of these processes are still not well understood. There are only a few dozen TDEs known to date (e.g. \citealt{vanVelzen20,vanVelzen21,Hammerstein23}), and a few hundred  confirmed CSAGNs (e.g. \citealt{MacLeod19,Graham20,LopezNavas22,LopezNavas23b}). Moreover, the identification of these events could be contaminated by circumnuclear supernovae (SNe; e.g. \citealt{Drake11}). \cite{Zabludoff21} proposed a comprehensive list of photometric and spectroscopic properties that may separate TDEs from classical AGNs and flaring AGNs, but these authors also note that no single TDE has been known to present all of the described features.

In this work, we present the discovery and follow-up of a $\sim 10^6 M_{\odot}$ BH candidate that started showing optical variations  in December 2019. The candidate was selected from the public ZTF alert stream by making use of the classifications provided by the ALeRCE (Automatic Learning for the Rapid Classification of Events) broker \citep{Forster21}. The source, SDSS J133519.91+072807.4 (hereafter SDSS1335+0728), presented a strong jump in its optical emission, and was denoted ZTF object ID `ZTF19acnskyy' from December 2019, but importantly exhibited no evidence of activity during the preceding two decades. From December 2019 to present, SDSS1335+0728 has shown stochastic optical variations. Based on our variability analysis and photometric and spectroscopic follow up, we discuss different scenarios that could explain the sudden flux jump and subsequent variability, and propose that an AGN in the process of turning on or a TDE in a $\sim 10^6 M_{\odot}$ BH could explain the observations. The paper is organised as follows. In Section \ref{section:selection} we present the identification of the nuclear variations in SDSS1335+0728. In Section \ref{section:data} we present the archival and follow-up data used to study the source. In Section \ref{section:analysis}, we present the variability, photometric, and spectroscopic analysis performed to understand the origin of the variations observed in SDSS1335+0728. Finally, in Section \ref{section:conclusion}, we conclude and summarise our findings.

\section{Identification of the nuclear transient ZTF19acnskyy in SDSS1335+0728}\label{section:selection}

The ALeRCE broker is actively processing the public ZTF alert stream (alerts are generated when a $5\sigma$ detection is obtained in the difference image of a source), providing light curve and stamp classifications of different variable and transient objects. The ALeRCE light curve classifier (LCC; \citealt{Sanchez-Saez21a}) uses 152 variability features computed from the ZTF alert stream, as well as average colours, to classify objects among 15 periodic, stochastic, and transient classes with high accuracy. Among the stochastic classes, there are three distinct AGN classes: core-dominated, host-dominated and blazars. 

During the development of the AleRCE LCC, we started to test its efficiency in identifying IMBHs and low-mass SMBH candidates from the ZTF alert stream. Specifically, we selected all the sources in the NASA-Sloan Atlas (NSA) catalogue\footnote{\url{http://nsatlas.org}} \texttt{v$_{1\_0\_1}$} \citep{Blanton11} with stellar mass (from K-correction fit for Sersic fluxes) below $10^{10} M_{\odot}$, and searched in the ALeRCE database for cross-matches to LCC objects with a classification of AGN (including the three AGN classes), using a radius of $1\farcs5$. We found 45 candidates, of which 12 were previously identified as IMBHs or low-mass SMBH candidates from Sloan Digital Sky Survey (SDSS; \citealt{York00}) spectroscopy \citep{Greene04,Greene07,Dong12,Chilingarian18,Liu18}. Among the remaining 33 sources, we found a very particular source, ZTF19acnskyy\footnote{SDSS1335+0728 was associated later in 2022 with the ZTF source ZTF22abyhaut, also detected in the nucleus of the galaxy.}, classified as core-dominated AGN by the LCC, and coincident with the nucleus of SDSS1335+0728 ($\alpha_{\rm J2000}:$ 203.833062\textdegree, $\delta_{\rm J2000}:$ 7.468791\textdegree, SDSS spectroscopic redshift 0.024, and stellar mass of $\sim 6\times10^9 M_{\odot}$ from the NSA catalogue), which started producing alerts (with stochastic variations) in the public ZTF stream since December 13th 2019 (MJD: 58830.567). SDSS1335+0728 was previously observed by ZTF, and it has observations in the ZTF data releases since March 2018 (MJD: 58198), but did not present $5\sigma$ detections in the ZTF difference images (and therefore did not produced alerts) until December 2019.

From the stellar mass of SDSS1335+0728 reported in the NSA catalogue, the predicted BH mass would be $\sim1.5\times10^{6} M_{\odot}$, assuming the scaling relations presented in Equations 4 and 5 of \cite{Reines15}, for local AGNs. SDSS DR8\footnote{\url{https://www.sdss4.org/dr17/spectro/galaxy_mpajhu/}} provide stellar velocity dispersions, stellar masses, and star formation rates (SFRs), obtained from the MPA-JHU catalogue (after the Max Planck Institute for Astrophysics and the Johns Hopkins University; \citealt{Kauffmann03,Brinchmann04,Tremonti04}). For SDSS1335+0728 the reported stellar velocity dispersion is 74.71 km/s. Using the  $M-\sigma$ relation of \citet{McConnell2013} for early type galaxies, we derive a BH mass of $\sim1.5\times10^{6} M_{\odot}$, consistent with the previous estimated value. While the reported stellar mass and SFR are $1.3\times10^{10} M_{\odot}$ and 0.228 $M_{\odot}/\text{yr}$, respectively. From this stellar mass, we can predict a BH mass of $3.3\times10^{6} M_{\odot}$. These predicted masses are right at the limit of what can be considered a low-mass SMBH, and are within the limits of BH masses expected for TDE hosts \citep{Wevers17}.

\begin{figure*}[htbp!]
\begin{center}
\begin{tabular}{cc}
\includegraphics[scale=0.5]{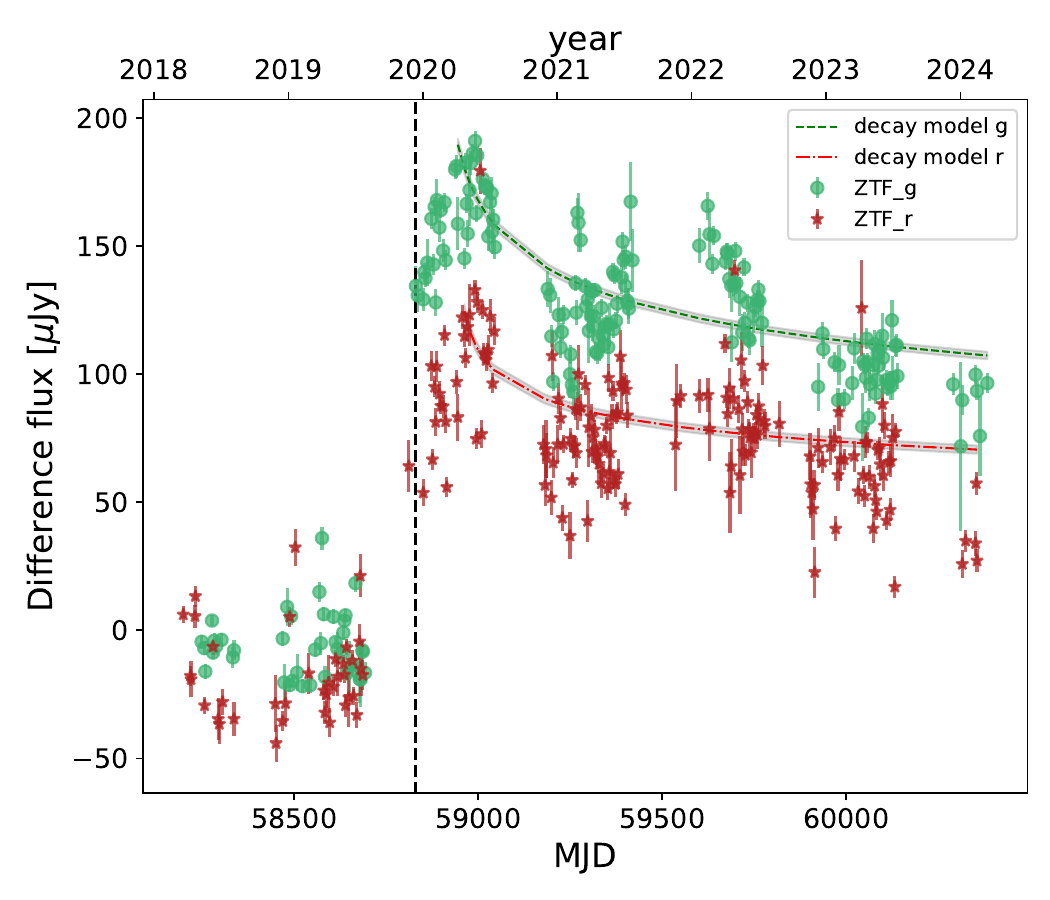} &
\includegraphics[scale=0.5]{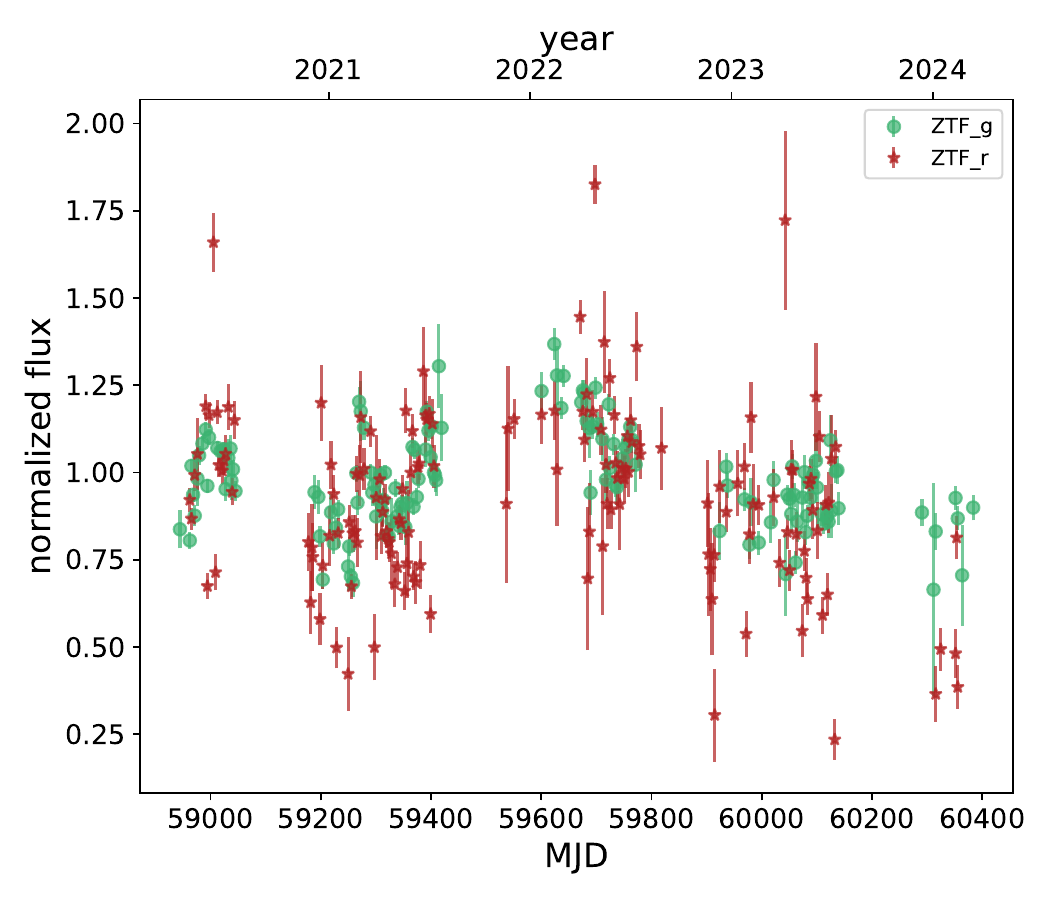} \\
\end{tabular}
\caption{ZTF light curves of SDSS1335+0728. \textit{Left:} ZTF forced photometry light curve at the position of SDSS1335+0728 in the $g$ (green circles) and $r$ (red stars) bands. The light curves have been binned using a bin of three days. The black dashed line marks the day when SDSS1335+0728 started triggering alerts in the ZTF alert stream. The green and red dashed lines show the results of the power-law decay fit (with their respective uncertainties shown in light grey). We obtain power-law indices $p$ of $-0.17\pm0.03$ and $-0.14\pm0.03$ in the $g$ and $r$ bands, respectively. \textit{Right:}  $g$ (green), and $r$ (red) band light curves obtained after normalising by the power-law decay fit.
 \label{figure:lc_dif}}
\end{center}
\end{figure*}

The left panel of Figure\,\ref{figure:lc_dif} shows the difference-flux light curve at the position of SDSS1335+0728, obtained by using the ZTF forced photometry service \citep{Masci23}, in the ZTF $g$ and $r$ bands. This corresponds to the point-spread-function (PSF)-fit flux measured over all the difference images of the target, which are constructed by subtracting a reference image from the single-epoch science images of the source. The figure shows binned light curves, 
constructed by measuring the median MJD and fluxes in bins of three days. For the errors we assumed that they combine as the root-mean-square\footnote{Following the implementation of Lightkurve v2.4. More details in \url{https://docs.lightkurve.org/index.html}}.

From the left panel of Figure\,\ref{figure:lc_dif}, we can see that prior to December 2019, the flux detected in the difference images was consistent with being non-variable (difference-flux around zero), since the first observation of the source in March 2018. In December 2019, SDSS1335+0728 increased its flux, reached its peak in MJD 58991 (May 2020), and has been slowly decaying since, showing stochastic variation for more than 1,550 days. The brightest absolute magnitudes (AB) observed in the difference images were $-16.9$, $-16.8$, and $-16.4$ in the ZTF $g$, $r$, and $i$ band, respectively.

Considering the temporal behaviour and the properties of SDSS1335+0728, we hypothesised two possible scenarios for its origin: (a) it corresponds to a $\sim 10^6 M_{\odot}$ BH that was either not accreting or was accreting at very low rates (a low-luminosity AGN; LLAGN; e.g. \citealt{Ho08,Ho09}) that turned on (i.e. a BH that increases its accretion rate enough as to form an accretion disc, potentially for the first time); or (b) it corresponds to a TDE (or another class of still unknown nuclear transient) in a $\sim 10^6 M_{\odot}$ BH.

The decay timescale and maximum brightness of ZTF19acnskyy are much slower and fainter than other known {classical} SNe events, respectively (e.g. \citealt{Perley20,Grayling23,Chen23}), and thus we rule out a classical SN to explain the variations observed in the nucleus of SDSS1335+0728. However, a potential origin for the behaviour of SDSS1335+0728 would be an exotic circumnuclear SN. iPTF14hls \citep{Arcavi17} and SN2018ibb \citep{Schulze24} are two examples of SNe that evolve extremely slowly compared to normal SNe events, with electromagnetic emission detected during $\sim 1000$ days, similar to what we observe in ZTF19acnskyy. However, the absolute magnitudes at the peak for iPTF14hls and SN2018ibb ($\sim-18.2$ and $-21.8$ in the $g$ band, respectively) are much brighter than that observed for ZTF19acnskyy, and they decayed more than five magnitudes after $\sim 1000$ days, while ZTF19acnskyy has only faded by $\sim0.82$ magnitudes in the $g$ band (in the difference-magnitude light curve), after 1550 days. Hence, we ruled out the hypothesis of a potential circumnuclear SN to explain the behaviour at the nucleus of SDSS1335+0728.

In case we confirm SDSS1335+0728 as a turning-on AGN, this source will allow us to observe an AGN as it becomes active, which will be extremely relevant for the understanding of the physics behind BH accretion, as well as AGN formation and evolution. Recently, \cite{Arevalo24} reported the discovery of new AGN activity in a former quiescent galaxy. However, the source was discovered several years after the transition event, and thus, it was not possible to follow up the source while the broad line region (BLR) was forming. Their work, though, provides an upper limit for the timescale of the appearance of broad emission lines (BELs) in AGN ignition events ($\sim$18 years for a $\sim 10^7 M_{\odot}$ BH). 

Transitions from quiescent states to active AGNs have also been observed in Low Ionization Nuclear Emission Line Regions (LINERs). For instance, \cite{Yan19} and \cite{Frederick19} reported the discovery of LINERs that transitioned into type 1 AGNs. SDSS1335+0728 is classified as composite (i.e. a galaxy whose emission lines could be explained by both AGN and/or SF star formation activity; also known as transition objects) in the MPA-JHU catalogue. Composite galaxies are at the limit of what can be considered a pure star-forming galaxy and galaxies with AGN activity \citep{Ho93,Veron97}. Their origin is still a matter of debate, although it has been proposed that some of them could host LLAGNs (e.g. \citealt{Ho08}). If we confirm the AGN nature of the variations in SDSS1335+0728, it will correspond to the first direct observation of a composite galaxy that transitioned into a classical AGN.

On the other hand, the faintest TDE known up to date corresponds to eRASSt J074426.3+291606 \citep{Malyali23},  with an absolute magnitude in the $g$-band peak of $-16.8$ (very close to the value observed for ZTF19acnskyy), and an exponential decay timescale of $\sim 120$ days in the optical range. Optically faint TDEs are normally observed to fade faster than their brighter counterparts \citep{vanVelzen20}, and thus \cite{Malyali23} proposed that eRASSt J074426.3+291606 corresponds to the first example of a potential new class of `faint and slow' TDEs. ZTF19acnskyy has been showing stable variations during more than 1550 days, and thus, if we confirm its TDE nature, this will correspond to the slowest and faintest optical TDE yet observed.

\section{Data collection for SDSS1335+0728}\label{section:data}

\subsection{Archival data}\label{section:archival_data}

To better understand the origin variations observed in ZTF19acnskyy and the properties of SDSS1335+0728, we collected single epoch UV/optical/infrared photometry for SDSS1335+0728 (including the whole galaxy), obtained prior to December 2019 from the NASA/IPAC Extragalactic Database (NED; \citealt{Helou91}), obtaining detections from the Galaxy Evolution Explorer (GALEX; \citealt{Martin05}; Kron flux in elliptical aperture), SDSS (Model flux), the Two Micron All Sky Survey (2MASS; \citealt{Skrutskie06}; extended object magnitude), and the Wide-field Infrared Survey Explorer (WISE; \citealt{Wright10,Mainzer14}; profile-fit/extended object magnitude). 

To investigate if SDSS1335+0728 presented previous variable activity, we searched for archival light curves from the following datasets: ZTF $g$, $r$, and $i$ bands (as mentioned previously); \textit{Gaia} DR3 \citep{GaiaDR3}, which includes data from July 2014 to May 2017; the Catalina Real-Time Transient Survey DR2 (CRTS; \citealt{Drake09}), including data from April 2005 to April 2016; the All-Sky Automated Survey for Supernovae (ASASSN; \citealt{Jayasinghe18}, using data from January 2012 to March 2024; and WISE, using observations from January 2014 to June 2023. 

In terms of spectroscopic observations, SDSS1335+0728 was observed prior to December 2019 by SDSS on February 2nd, 2007 (MJD 54156), using a $3''$ fibre (R=1500 at 3800 $\AA$, R=2500 at 9000 $\AA$). More recently, the source was observed by the Large Sky Area Multi-Object Fiber Spectroscopic Telescope (LAMOST; \citealt{Yan22}) on May 12th, 2015 (MJD 57155) using a $3\farcs5$ fibre (R=1800 at 5500 $\AA$).

We did not find archival radio or X-ray detections for SDSS1335+0728. The source remained undetected by the Very Large Array Sky Survey in April 2019 and November 2021 (VLASS, RMS is 0.13 mJy/beam), and in the Rapid ASKAP Continuum Survey (RACS, RMS is 0.2 mJy/beam) in April 2019. Moreover, it was not detected by the German eROSITA (extended ROentgen Survey with an Imaging Telescope Array; \citealt{Merloni12,Predehl21}) All-Sky Survey (eRASS), including the five eRASS available observations (MJDs: 58848, 59030, 59214, 59399, and 59583) and the stacked images of the first four eRASSes. 

\subsection{Follow-up data}\label{section:follow_data}

In mid-2021, we started a follow-up campaign to further understand the origin of the variation in the nucleus of SDSS1335+0728. In July 2021 (MJD 59426), July 2022 (MJD 59775), July 2023 (MJD 60139), and February 2024 (MJD 60344), we requested target of opportunity (ToO) observations of SDSS1335+0728  with \emph{Swift} \citep{Burrows05}, observing the source with exposure times of 3, 4.6, 1.1, and 2 ksec, respectively. The data reduction was performed following standard routines as described by the UK \emph{Swift} Science Data Centre (UKSSDC), using the software in HEASoft version 6.30. In addition, a \emph{Chandra} \citep{Weisskopf00} Director’s Discretionary Time (DDT) observation was obtained in April 2024 (MJD 60403). The data reduction was performed using CIAO v4.16 and CALDB v4.11.0.

Moreover, we performed a spectroscopic follow-up of SDSS1335+0728. First, in June 2021 (MJD 59372) and July 2021 (MJD 59396), we obtained optical spectra with the Goodman spectrograph at SOAR \citep{Clemens04}, using the Red Camera with the 400M2 (total exposure time of 1800 seconds; wavelength range $5100 \AA - 8950 \AA$) and 400M1 (total exposure time of 3000 seconds; wavelength range $3300 \AA - 7030 \AA$) configurations, respectively. In both observations, we used the $1''$ slit, with $2\times2$ binning (R$\sim1270$). The spectra were reduced using our own custom IRAF routines. Since the source did not vary drastically in the optical range between MJD 59372 and MJD 59396, we combined both observations, to improve the wavelength coverage. The signal-to-noise per resolution element below $3800 \AA$ was $\sim2$, and thus we cut the spectrum at this wavelength.

In July 2022 (MJD 59789) we obtained 1.2 hours of DDT (programme ID 109.24F5.001) to observe SDSS1335+0728 with VLT/X-shooter \citep{Vernet11}, using the following configuration: a) UBV arm with 0\farcs5 slit (R=9861), $1\times1$ binning, and total exposure time of 1500 seconds, b) VIS arm with 0\farcs4 slit (R=18340), $1\times1$ binning, and total exposure time of 1050 seconds, and c) NIR arm with 0\farcs4 slit (R=11424), $1\times1$ binning, and total exposure time of 1500 seconds. The observations were done using an OSSO pattern, with a fixed RA offset of $0''$ and a Fixed DEC offset of $20''$. The data were reduced using the standard ESO pipelines (\citealt{Modigliani10}; version 3.6.3.1) within the Reflex \citep{Freudling13} front end.\footnote{\url{https://ftp.eso.org/pub/dfs/pipelines/instruments/X-shooter/xshoo-reflex-tutorial-3.6.3.pdf}} Unfortunately, the NIR observation had an overall signal-to-noise of 4.9 and was dominated by telluric lines, and thus we did not use the NIR arm in this work. For the case of the UVB arm, the signal-to-noise at wavelengths shorter than 3800 $\AA$ was $\lesssim5$, and therefore, we cut this arm at 3800 $\AA$. In the VIS arm, we cut the spectrum at 5500 $\AA$ (where the signal-to-noise was $\lesssim3$). 

In July 2023 (MJD 60137), we obtained observations with the Low Resolution Imaging Spectrometer (LRIS) at Keck I \citep{Oke95}, with the 600/4000 Blue grism ($2\times2$ binning, R=3495), and the 600/7500 Red grating ($2\times1$ binning, R=8835). We obtained two observations, 300 second exposures each, with 1\farcs5 and 0\farcs7 slits, respectively. The two adopted slit widths will later allow us to compare aperture loss issues. The data were reduced using the standard Keck \textsc{lpipe} pipeline.\footnote{\url{https://sites.astro.caltech.edu/~dperley/programs/lris/manual.html}}.

Finally, in January 2024 (MJD 60337), we obtained a new spectrum with the Goodman spectrograph using the Red Camera with 400M1 (total exposure time of 3000 seconds). As for the 2021 observation, the spectrum was reduced using IRAF. A summary of all the available spectroscopic observations is presented in Table \ref{table:spec_sum}.

In the following sections, we use all the available photometric and spectroscopic data to disentangle the nature of the variations currently observed in SDSS1335+0728.

\begin{table*}[htpb]
  \begin{center}
    \caption{Spectroscopic observations of SDSS1335+0728.}
    \label{table:spec_sum}
    \begin{tabular}{ccccc} 
   
\hline
\hline

Instrument & MJD & Aperture & Seeing & Resolution (R)  \\

\hline

SDSS &  54156 & 3$''$ & 1\farcs64 & 1500 at 3800 $\AA$, 2500 at 9000 $\AA$ \\
LAMOST & 57155 & 3\farcs5 & 4$''$ & 1800 at 5500 $\AA$ \\
\hline
SOAR/Goodman & 59396 & 1$''$ & 1\farcs5 & $\sim1270$ \\ 
VLT/X-shooter & 59789 & 0\farcs4 (0\farcs5 UBV) & 0\farcs7 & 9861 in the UBV arm and 18340 in the VIS arm \\
Keck/LRIS & 60137 & 0\farcs7 and 1\farcs5 & 0\farcs45 &  3495 in the Blue grism and 8835 in the Red grating\\ 
SOAR/Goodman & 60337 & 1$''$ & 1\farcs3 & $\sim1270$ \\

\hline
\hline

  \end{tabular}
  \end{center}
\end{table*}

\section{Understanding the nature of the variations in the nucleus of SDSS1335+0728}\label{section:analysis}

\subsection{Optical variability}\label{section:opt_var}

We used the ZTF optical light curves to explore the potential for ZTF19acnskyy to be a TDE in the nucleus of SDSS1335+0728. We fitted the ZTF $g$ and $r$ difference light curves shown in the left panel of Figure\,\ref{figure:lc_dif} using the decaying power-law model presented in Equation 3\footnote{$L_{\text{TDE}}(t)\propto[(t-t_{\text{peak}}+t_0)]^p $ for $t>t_{\text{peak}}$.} of \cite{vanVelzen21}, which has been successfully used to fit TDE light curves. TDEs are predicted to have light curves with a smooth power-law decay with $p$ index of $-5/3$ \citep{Rees88,Phinney89}, although some deviations from this value have been observed ($-3.82 \lesssim p \lesssim -0.78$; \citealt{Hammerstein23}). For the case of ZTF19acnskyy, we obtain much flatter power-law indices $p$ of $-0.17\pm0.03$ and  $-0.14\pm0.03$ in the ZTF $g$ and $r$ bands respectively. These fits are shown in the left panel of Figure \ref{figure:lc_dif} (dashed lines). 

The right panel of Figure \ref{figure:lc_dif} shows the ZTF difference light curves, but normalising them by the power-law decay fit obtained for each band. From this figure we can see that the variations of ZTF19acnskyy differ from the smooth power-law decay expected for TDEs, showing stochastic variations similar to those observed in classical AGNs. The variations observed in both bands are consistent, although the $r$-band light curve shows more scatter due to its lower signal-to-noise. The excess variance $\sigma^2_{\text{rms}}$\footnote{$\sigma^2_{\text{rms}}=(\sigma_{LC}^2-\overline{\sigma}_{x}^2)/\overline{x}^2$, with $\sigma_{LC}$ being the standard deviation of the light curve, $\overline{\sigma}_{x}$ the average photometric error, and $\overline{x}$ the average magnitude or flux (\citealt{Sanchez17} and references therein).} of the normalised flux light curve in the $g$ band is 0.017, while for the original difference-flux light curve (using epochs after MJD 58830) is 0.039. Both values are in agreement with values measured for classical AGNs \citep{Simm16}.

\begin{figure*}
\sidecaption
  \includegraphics[width=12cm]{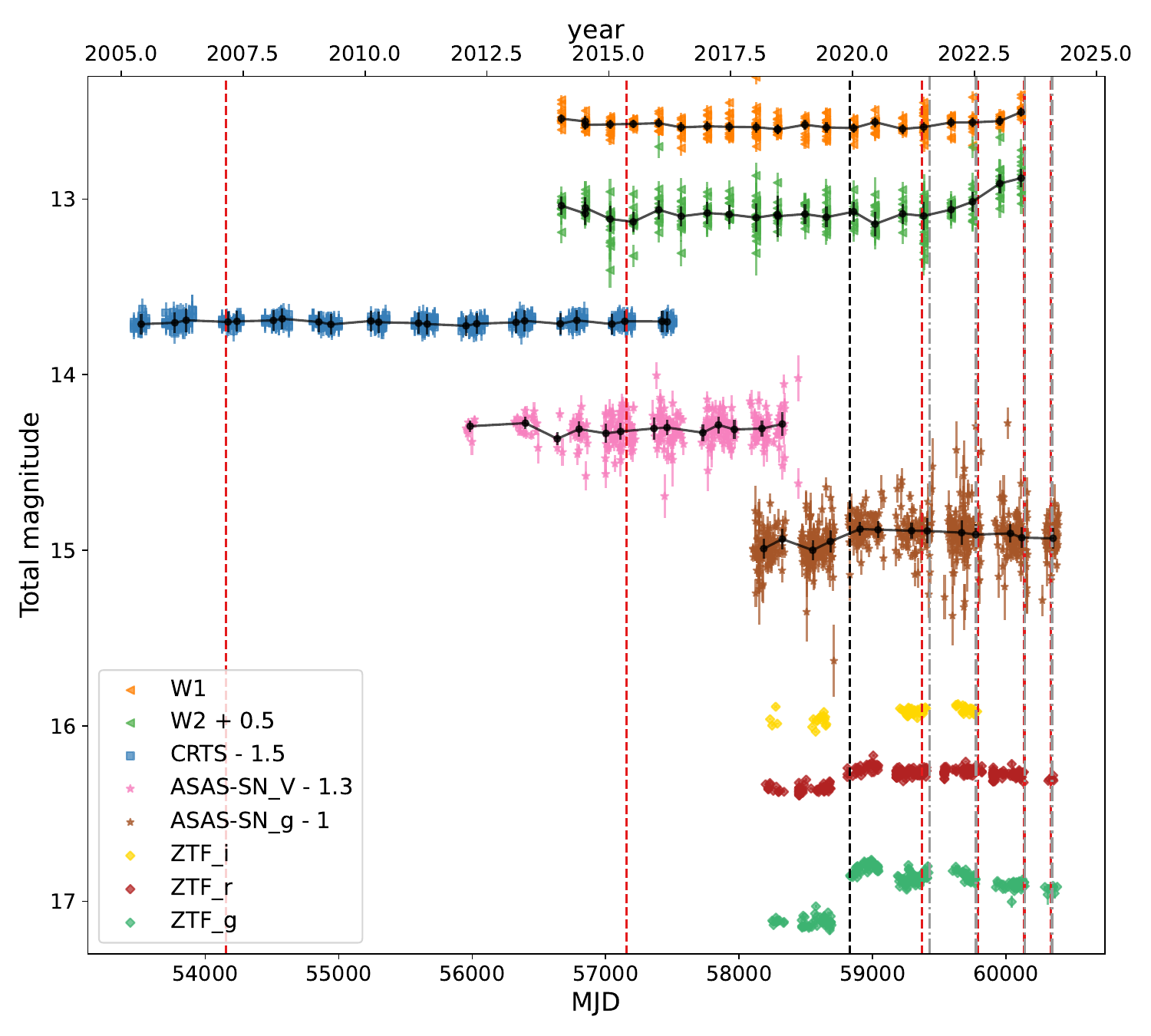}
   \caption{Total magnitude light curves of SDSS1335+0728 in the ZTF $g$ (green), $r$ (red), and $i$ (yellow) bands, CRTS (blue), WISE W1 (gold), and W2 (light green) bands, and in the ASAS-SN V (pink) and $g$ (brown) bands. The black circles show the binned light curves constructed using bins of 180 days. The black dashed line marks the day when ZTF19acnskyy started triggering alerts in the ZTF alert stream. The red dashed lines mark the dates with available spectra, and the grey dashed-dotted lines show the dates with available \emph{Swift}/UVOT observations (more details on Section \ref{section:data}). \label{figure:lc_all}}
\end{figure*}

Figure\,\ref{figure:lc_all} shows the ZTF Forced Photometry total magnitude light curve of SDSS1335+0728, obtained by correcting the difference-flux by the total flux measured on its respective reference image. It can be seen that the optical flux increased in all the ZTF bands in December 2019, with the largest variation being observed in the ZTF $g$ band. After this increase in luminosity, the source has shown stochastic variations (according to the LCC top-level classification) for more than 1550 days ($\sim4.2$ years), but its flux has declined by a factor of $\sim2$ since its peak. 




To investigate if SDSS1335+0728 presented previous optical variability, we used the archival CRTS and ASSASN light curves. These light curves are shown in Figure\,\ref{figure:lc_all}. We did not use the \textit{Gaia} DR3 data since the source is classified as non-variable, and thus its time series is not currently available. These optical light curves cover almost 19 years of data, from April 2005 until March 2024, but have low signal-to-noise ratios, and therefore we binned them to have a better idea of any possible variations. We used the same procedure described in Section \ref{section:selection}. By visual inspection, we see that the source did not vary in the optical range before December 2019.  The ASASSN\_V band light curve shows modest evidence of an increase in the flux since December 2019, especially after binning the light curve, although this light curve remains quite noisy. 

To constrain quantitatively the variability properties of the archival optical light curves, following \cite{Sanchez17}, we calculated the probability that the source is intrinsically variable ($P_{\text{var}}$; see \citealt{Paolillo04} and references therein), and the excess variance $\sigma^2_{\text{rms}}$ (in magnitudes), and classified as variable the light curves showing $P_{\text{var}}>0.95$ and $\sigma^2_{\text{rms}}-err(\sigma^2_{rms})>0$. For this, we used the binned archival light curves shown in Figure\,\ref{figure:lc_all}. None of the CRTS and ASSASN light curves is classified as variable according to the criteria, with all having $P_{\text{var}}\lesssim0.5$ and  $\sigma^2_{\text{rms}}-err(\sigma^2_{rms})<0$. On the other hand, all the ZTF light curves are classified as variable. Consequently, we can confirm that SDSS1335+0728 has not varied in the optical range in the last $\sim2$ decades until it started showing variations in December 2019.

From the optical variability, we conclude that the turning-on AGN scenario is plausible, while the TDE scenario is less probable, especially considering the slow temporal evolution of ZTF19acnskyy. For the case of a potential TDE in an AGN, simulations still predict a smooth power-law decay (e.g. \citealt{Chan20}), and available observations show that the timescales of these events are similar to those of classical TDEs (e.g. \citealp{Merloni15,Ricci20,Homan23}). However, given the current declining trend of the optical light curve, it is still too soon to rule out a transient event. Future observations from ZTF and LSST will be key to further understanding the variability observed in SDSS1335+0728. 

\subsection{Infrared variability}\label{section:mir_var}

To study the mid-infrared evolution of SDSS1335+0728, we used the archival WISE light curve, which is shown in Figure\,\ref{figure:lc_all}. The W1 and W2 light curves show a slow increase in their flux since $\sim$ June 2022 (i.e. $\sim900$ days since the first ZTF alert), which is more evident in the W2 band, where the flux increased more than two times in $\sim$one year (between June 2022 and June 2023).

As for the optical light curves (see Section \ref{section:opt_var}), we binned the WISE light curves using the procedure presented in Section \ref{section:selection} and measured the $P_{\text{var}}$ and $\sigma^2_{\text{rms}}$ from them. When using the full timespan of the WISE light curves, for both bands, we obtain $P_{\text{var}}<0.95$ and $\sigma^2_{\text{rms}}-err(\sigma^2_{rms})<0$. However, these full light curves are dominated by the observations prior to 2022. If we only include observations obtained after the first ZTF alert, the W2 light curve is classified as variable.  

\begin{figure}[htbp!]
\begin{center}

  \includegraphics[width=\linewidth]{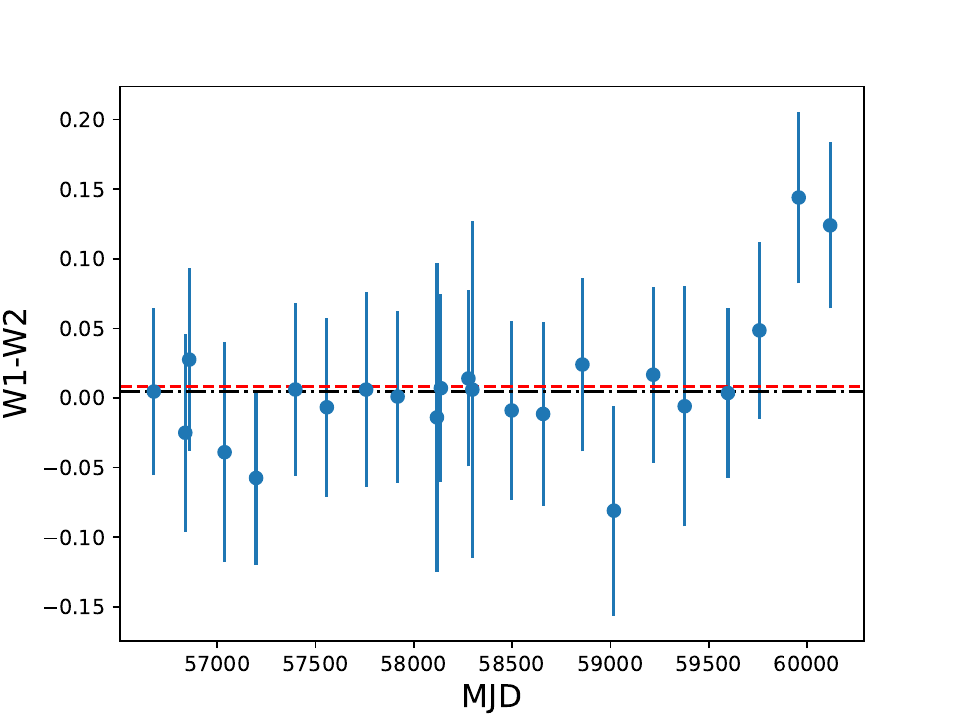} 

\caption{W1-W2 WISE colour evolution for SDSS1335+0728. The red and black dashed lines show the mean and median W1-W2 colour, respectively. \label{figure:wise_col}}
\end{center}
\end{figure}

We show in Figure \ref{figure:wise_col} the W1-W2 WISE colour evolution of SDSS1335+0728. From the figure, we can see that the W1-W2 colour changed from $\sim 0.0$ to $\sim 0.14$ at MJD 59958 (1128 days after the first ZTF alert).

The delayed response of the MIR emission is much larger than what is expected from the UV/optical emission reprocessing by a potential preexisting dusty torus around the SMBH (i.e. infrared dust echo). For classical AGNs, from previous dust reverberation mapping studies, we know that delays of the order of 900 days (or 879 days in rest frame for SDSS1335+0728) in the 3.6--4.5 $\mu$m wavelength range are expected for AGNs with bolometric luminosities larger than $10^{46}$ erg/s \citep{Lyu19,Yang20RM}. For instance, by using equation 16 from \cite{Lyu19}, a time delay of 879 days in W2, implies a bolometric luminosity of $\sim2.4\times10^{46}$ erg/s, which is much brighter than the nuclear emission observed from SDSS1335+0728 (see Section \ref{sec:sed}). Moreover, TDEs with known MIR variations have shown much shorter MIR lags (e.g. \citealt{vanVelzen16,Jiang19,vanVelzen21b}), with typical estimated timescales of a few months, for the reprocessing of the X-ray/UV/optical signal by the dusty torus.  

Evolution towards redder MIR colours has been observed before in CSAGNs (e.g. \citealt{Lyu22,LopezNavas23b} and references therein) and in the newborn AGN reported by \cite{Arevalo24}, and it has been associated with reprocessed emission from the AGN dusty torus. Classical AGNs have values of W1-W2 larger than 0.6 \citep{Wright10,Comparat20}, while values lower than 0.4 are normally associated with quiescent or star-forming galaxies. However, low-luminosity AGNs (LLAGNs) can have W1-W2 colours of $\sim 0.08$ \citep{Lyu22}, as the host galaxy dominates their MIR emission. For the case of SDSS1335+0728, the recent MIR flux and colour evolution could indicate that a new dusty torus is just being formed in an LLAGN, {or that the dust reprocessing is happening in a much larger dusty structure than the ones typically observed in classical AGNs or TDEs, with a size of few light years}.  

\subsection{Ultraviolet photometry}\label{section:uv_analysis}

GALEX detected SDSS1335+0728 on April 16, 2004 (MJD 53111) in both NUV and FUV filter bands. Around 17 years later, we started observing the source with \emph{Swift}/UVOT in the UVW1, UVW2, UVM2, U, B, and V bands. The {\sc uvotsource} task was used to perform aperture photometry on the \emph{Swift}/UVOT images, using a circular aperture radius of $5''$ centred on the source. We excluded from the analysis the \emph{Swift}/UVOT observation from MJD 60344, since the UVW1, UVW2, and UVM2 images presented tracking issues.

\begin{figure}[tb]
    \centering
    \includegraphics[width=\linewidth]{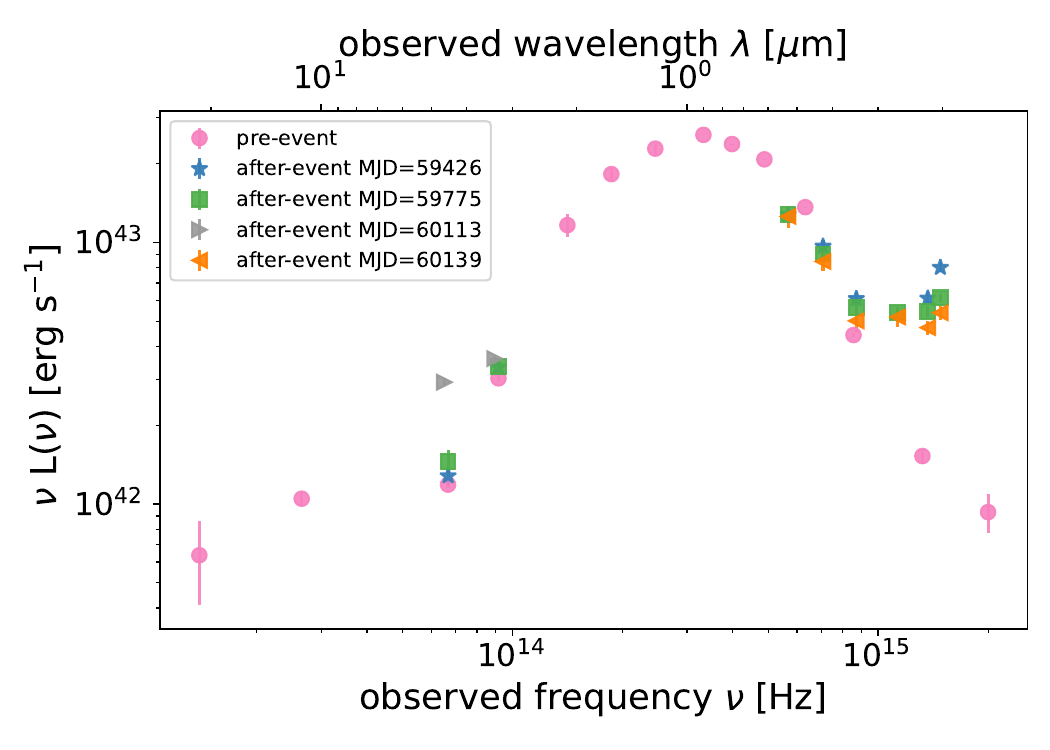}
\caption{Spectral energy distribution of SDSS1335+0728 before December 2019 (pink circles) constructed using data from WISE, 2MASS, SDSS, and GALEX; and at MJD 59426 (blue stars), MJD 59775 (green squares), MJD 60113 (grey triangles), and MJD 60139 (orange triangles) constructed using data from WISE and \emph{Swift}/UVOT. \label{figure:sed}}
\end{figure}

We used the available photometry described in Section \ref{section:data}, obtained prior to and after the event (excluding the MJD 60344 \emph{Swift}/UVOT observation), to construct a spectral energy distribution (SED) of SDSS1335+0728, which is shown in Figure\,\ref{figure:sed}. We show the four \emph{Swift}/UVOT measurements independently and complement them with contemporaneous WISE data, except for the MJD 60139 observation, for which there is no contemporaneous data publicly available (but we show the most recent WISE observation).

From Figure \ref{figure:sed}, we can see that the UV flux increased by a factor of $\sim$4 between archival GALEX observations (2004) and the first \emph{Swift}/UVOT measurement (2021). It is worth remarking that this difference is still robust even if one takes into account the difference in the filters involved. 

The \emph{Swift}/UVOT measurements show that SDSS1335+0728 has been varying in the UV since its outburst, with the highest flux being detected in the July 2021 observation and the lowest one in the July 2023 observation. In the UVW2 band the source varied 0.43 magnitudes between the 2021 and the 2023 observations, while in the UVM2, the source varied 0.28 magnitudes in the same period. The recent UV variability exhibits a bluer when brighter behaviour, with observed \emph{Swift} colour UVW2-UVM2 of $-0.208$, $-0.047$, and $-0.057$, and UVM2-U of 0.484, 0.527, and 0.554, in July 2021, July 2022, and July 2023, respectively. In comparison, the $g-r$ optical colour measured from the difference forced photometry has mean, median, and standard deviation values of $-0.54$, $-0.52$, and 0.25, respectively. This difference in the behaviour of the UV and optical colours of SDSS1335+0728 could indicate that a large fraction of the transient event energy is being released in the UV (presumably near to the SMBH).

These results support the hypothesis of a turning on BH in the nucleus of SDSS1335+0728; as a bright UV emission and a bluer when brighter behaviour is normally observed in accretion discs of classical AGNs, while for the case of TDEs, a rather constant colour is expected \citep{Zabludoff21}.

\subsection{X-ray photometry and spectroscopy}\label{section:xray_analysis}

As mentioned in Section \ref{section:archival_data}, eROSITA did not detect SDSS1335+0728 in any of the five available eRASS epochs in the same field of the source. We note that the first epoch was obtained on December 31, 2019, after the first ZTF alert. We obtained eROSITA upper limits  from the stacked images \citep{Tubin24}, which are calculated using X-ray photometry on the eROSITA standard calibration data products (counts image, background image, and exposure time), following the Bayesian approach described by \citet{Kraft91}. We consider a circular aperture with a radius given by a PSF encircled energy fraction of EEF = 0.75 ($\sim 30\arcsec$) and computed upper limits with a confidence interval of 99.87\% (this corresponds to a one-sided $3\sigma$ level). To calculate energy conversion factors (ECFs), we assumed an absorbed power-law model with spectral index $\Gamma=2$ and Galactic absorption $3\times$10${^{20}}$ cm$^{-2}$. We computed upper limits for the most sensitive eROSITA band with $0.2-2.3$~keV, as well as in nominal soft (0.2--0.6 keV), medium (0.6--2.3 keV), hard ( 2.3--5.0 keV), and total (0.2--5.0 keV) bands, obtaining values of $3.98\times10^{-14}$, $3.09\times10^{-14}$, $3.06\times10^{-14}$, $2.16\times10^{-13}$, and $4.00\times10^{-14}$ erg s$^{-1}$ cm$^{-2}$, respectively. From this, we estimate a $3\sigma$ upper limit in the luminosity of SDSS1335+0728 in the $0.2-2.3$~keV energy band of $5.17\times10^{40}$ erg s$^{-1}$.  

Additionally, in the first three \emph{Swift} observations (MJD 59426, 59775, and 60139), the source was not detected by the \emph{Swift}/XRT instrument. We used the longest observation (4.6 ksec in MJD 59775) to get an idea of the potential 0.5-10 keV band flux limit. We used WebPIMMS\footnote{https://heasarc.gsfc.nasa.gov/cgi-bin/Tools/w3pimms/w3pimms.pl} and estimated a count-rate of 1.3$\pm$0.5 $\times$10${^{-3}}$ cts $\hspace{0.1cm}$ s$^{-1}$ from a circular region of 20$''$ centred on the source. Then, assuming a power-law model with spectral index $\Gamma=2$ and Galactic absorption $3\times$10${^{20}}$ cm$^{-2}$,  we obtain estimated flux limits of $4.76\times$10$^{-14}$ erg s$^{-1}$ cm$^{-2}$ in the $0.5-10$~keV band, and $3.08\times$10$^{-14}$ erg s$^{-1}$ cm$^{-2}$ in the $0.2-2.3$~keV band, consistent with the upper limits provided by the eROSITA observations. If we instead use a black body model with kT=100 eV, we obtain estimated flux limits of $3.11\times$10$^{-14}$ erg s$^{-1}$ cm$^{-2}$ in the $0.5-10$~keV band, and $7.74\times$10$^{-14}$ erg s$^{-1}$ cm$^{-2}$ in the $0.2-2.3$~keV band. 

\begin{figure}[tb]
    \centering
    \includegraphics[width=\linewidth]{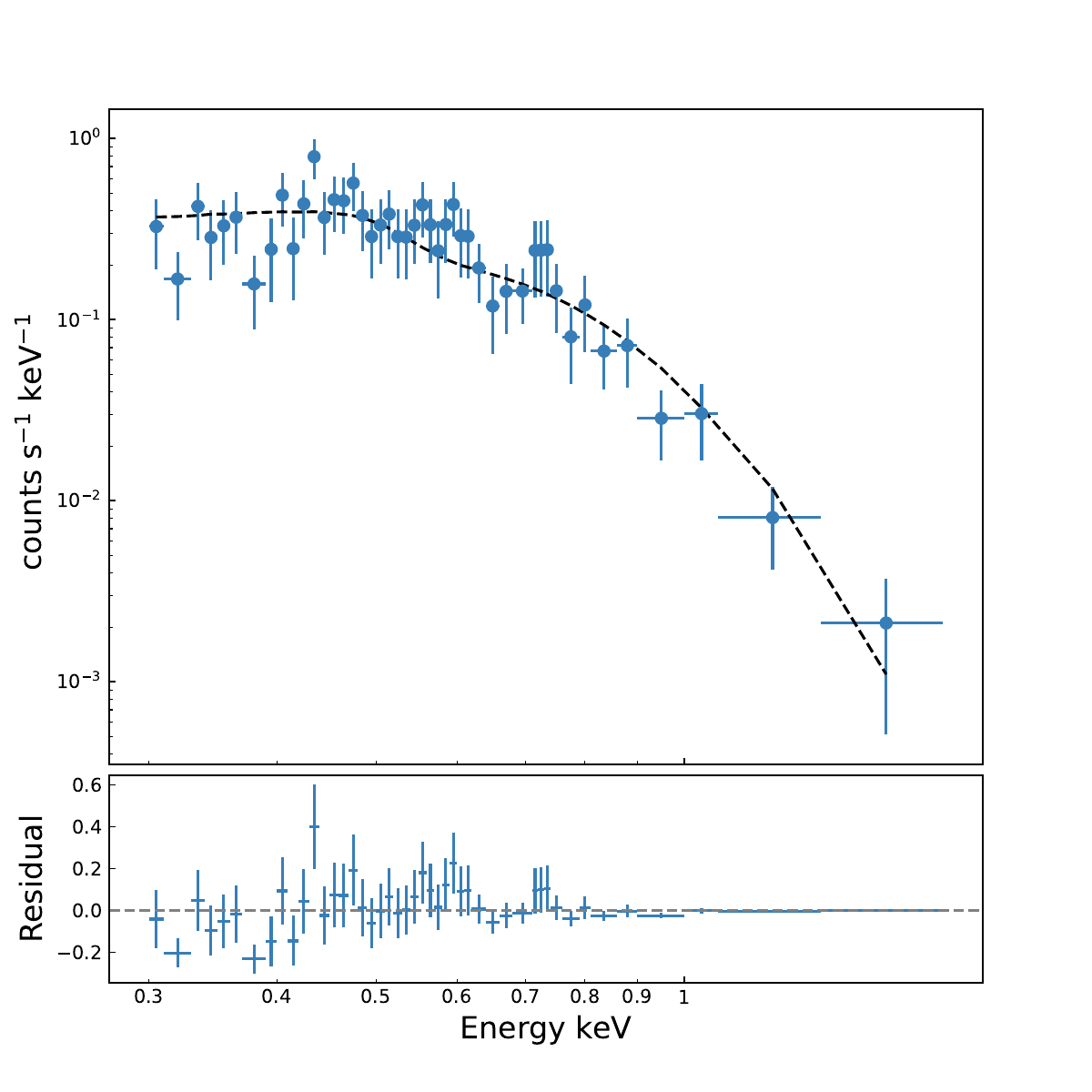}
\caption{\emph{Swift}/XRT spectrum of the source detected at a distance of 6\farcs0$\pm$3\farcs6 from the nucleus of SDSS1335+0728, obtained on February 4 2024. \textit{Top:} The blue circles show the data, and the black dashed line shows the black body model with kT$=104 \pm 9$ eV. \textit{Bottom:} Residual of the modelling (the difference between the original spectrum and the fit).  \label{figure:swift_spec}}
\end{figure}

Notably, on February 4 2024 (1514 days after the first ZTF alert), a source was detected in the 2 ksec \emph{Swift}/XRT observation at a distance of 6\farcs0$\pm$3\farcs6 from the nucleus of SDSS1335+0728\footnote{This detection was also reported in \citep{2024ATel16576}}. To extract the spectrum, we used the {\sc swifttools} v3.0, following the prescriptions of \cite{Evans09}. The obtained spectrum shows very soft emission, which is shown in Figure \ref{figure:swift_spec}. We modelled the spectrum with a black body model with a temperature of kT$=104 \pm 9$ eV. We obtained a flux of $7.4\times$10$^{-12}$ erg s$^{-1}$ cm$^{-2}$ in the $0.3-2$~keV band, which is two orders of magnitude larger than the eROSITA and \emph{Swift} upper limits for previous epochs. Using the redshift of the host galaxy, we obtain a luminosity of $9.62\times$10$^{42}$ erg s$^{-1}$ in the $0.3-2$~keV band.

\begin{figure}[tb]
    \centering
    \includegraphics[width=\linewidth]{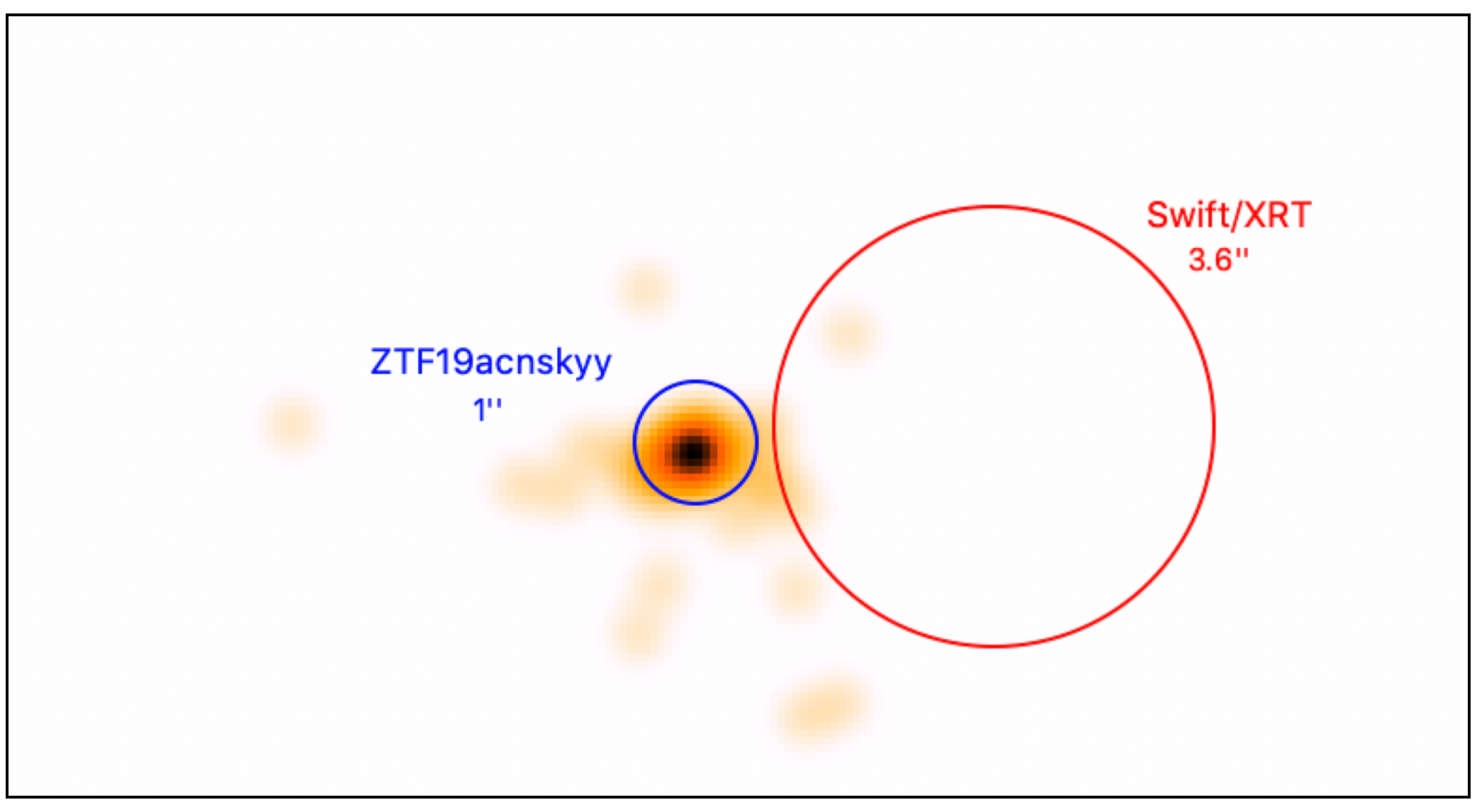}
\caption{\emph{Chandra} image of SDSS1335+0728 in the $0.3-2$~keV energy band, obtained on April 3 2024. The blue circle (with a radius of 1$''$) shows the ZTF alert position (ZTF19acnskyy), and the red circle (with a 3\farcs6 radius) is centred on the centroid of the first \emph{Swift}/XRT detection. \label{figure:chandra}}
\end{figure}

\emph{Swift} entered in safe mode on March 15 2024, due to the failure of a gyroscope, and therefore we could not confirm if the X-ray detection was nuclear or off-nuclear. To confirm the position of the X-ray emission, we requested a DDT observation with \emph{Chandra}. The source was observed on April 3 2024, during 2 ksec. Figure \ref{figure:chandra} shows the \emph{Chandra} image in the $0.3-2$~keV energy band, highlighting the position of the ZTF alert (ZTF19acnskyy) and the position of the first \emph{Swift}/XRT detection. The {\sc csmooth} task included in CIAO was used to adaptatively smooth the image, using a fast Fourier transform algorithm and a minimum and maximum significance level of the signal-to-noise of 3 and 4, respectively. The image shows a source located at RA: 13:35:19.93 DEC: +7:28:07.48, that is, coincident with the nucleus of SDSS1335+0728 and the position of ZTF19acnskyy. From this, we confirm that the previous \emph{Swift}/XRT detection comes from the nucleus of the source. We used the routine {\sc specextract} to extract the spectral products, using circles of 2$''$ and 10$''$ radii for the source and background, respectively. We modelled the spectrum with a black body model with kT$=80 \pm 7$ eV, obtaining a flux of $9.7\times$10$^{-12}$ erg s$^{-1}$ cm$^{-2}$ in the $0.3-2$~keV band.

The \emph{Swift}/XRT spectrum shows a shape similar to that observed in 1ES 1927+654 while its X-ray corona was being reformed \citep{Ricci20}, which could indicate that an AGN corona has started to form in the nucleus of SDSS1335+0728. Moreover, the spectra of X-ray-detected TDEs tend to show soft black body-like emission and temperatures of kT $\sim 0.04 - 0.12$ keV \citep{Zabludoff21,Saxton21}, and thus, the recent \emph{Swift}/XRT detection could point to a TDE origin. Delayed X-ray brightening has been observed in optically detected TDEs, but the observed timescales for these events are a few hundred days \citep{Guolo24,Wevers24}, thus much shorter than the ones observed for SDSS1335+0728. We have initiated a more intensive follow-up campaign with \emph{Swift}/XRT observations to confirm which of these hypotheses is correct. The analysis of this campaign, including a more detailed assessment of the \emph{Chandra} observation, will be presented in a forthcoming paper.

\subsection{Modelling the spectral energy distribution }\label{sec:sed}

To better understand the properties of the galaxy SDSS1335+0728 prior to and after December 2019, we modelled the SED presented in Figure \ref{figure:sed} using the Bayesian Markov chain Monte Carlo based code AGN\texttt{FITTER} \citep{Calistro-Rivera16}. This code models the SED of AGNs by combining multiwavelength models of the host-galaxy and AGN emission. In particular, four components are included: AGN accretion disc emission (big blue bump or BBB), the hot dust emission from the AGN torus, the stellar population of the host-galaxy, and the emission from the cold dust in galactic star-forming regions. The model returns the following parameters for the host-galaxy: stellar mass ($M_{\star}$), age, SFR, star-formation history timescale (SFH $\tau$), galaxy reddening [$E(B-V)_{\text{gal}}$], among others. For the AGN component, the code returns the torus and the BBB emission. From this, we can obtain integrated luminosities for the accretion disc and the torus. We run AGN\texttt{FITTER} using 10,000 walkers. Following \cite{Calistro-Rivera21}, we sampled each parameter from the model using 100 random realisations from the posterior, and we measured the luminosity of the accretion disc [$L_{\text{BBB}}$(0.05-1)] by integrating the BBB emission in the wavelength rage $0.05 - 1 \mu$m.

We first used AGN\texttt{FITTER} to model the SED of SDSS1335+0728 prior to December 2019 (preSED-model), using the photometry obtained from  WISE, 2MASS, SDSS, and GALEX. Then, to model the SED after December 2019 (afterSED-model), we combined the first \emph{Swift}/UVOT observation obtained in July 2021 (MJD 59426), with the photometry from 2MASS and WISE obtained prior to December 2019, assuming that the source did not vary in the infrared range up to that date. This allowed us to cover a larger range in wavelength. We assume throughout that the host-galaxy properties are constant, and fixed the stellar age, SFH $\tau$, and $E(B-V)_{\text{gal}}$ parameters, using the output of the preSED-model within $1\sigma$. The most recent WISE observations show a new MIR component, which could be associated with dust echo (torus reprocessing). Unfortunately, we do not have access to recent near-infrared (NIR) observations, and we cannot assume anymore that the source does not vary in this range. Thus, we cannot properly model an SED that includes this new component. Therefore, we just modelled the SED for the July 2021 observation.

\begin{figure*}
\sidecaption
\includegraphics[width=12cm]{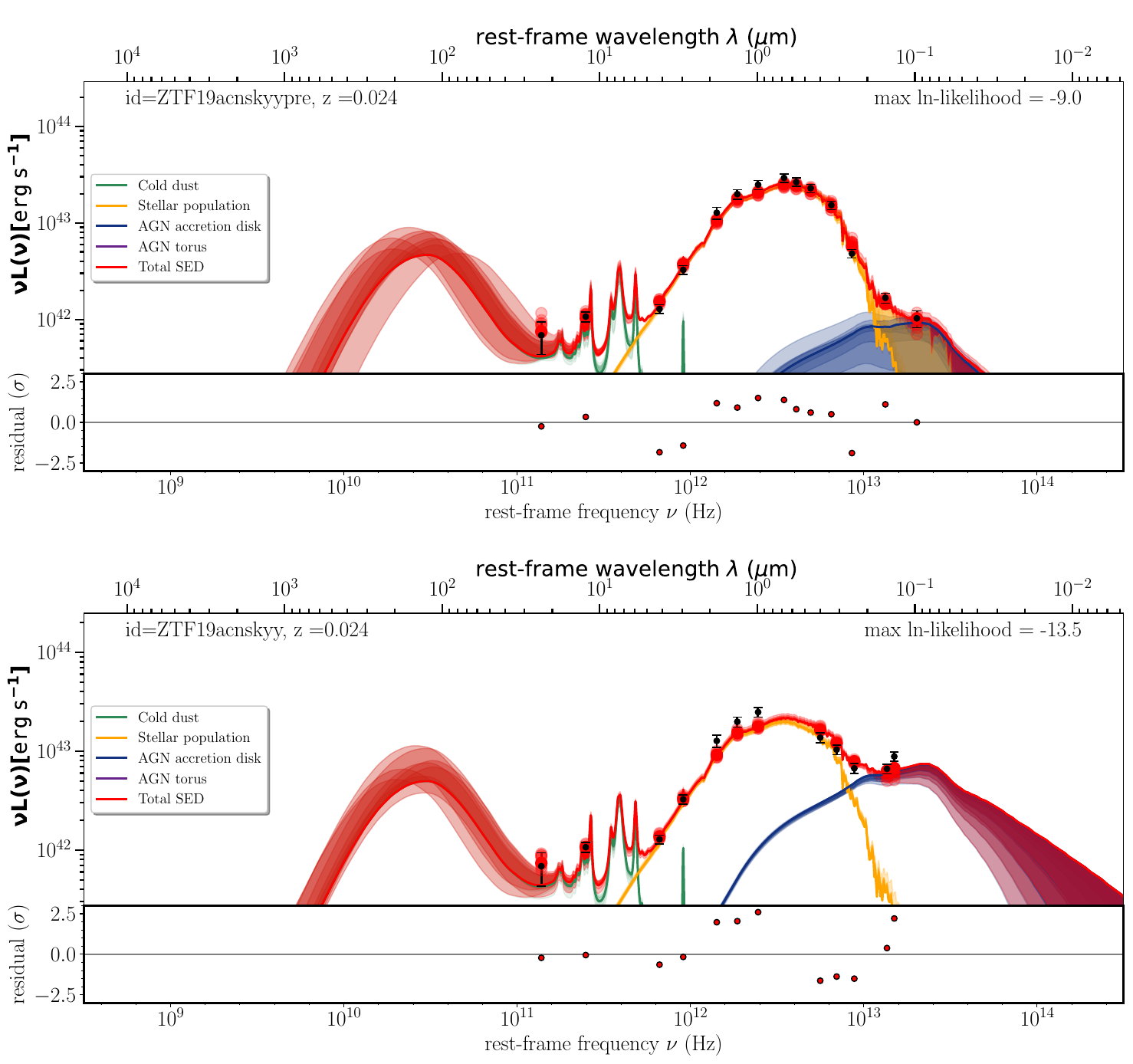}
\caption{Modelling of the SED of SDSS1335+0728 obtained using AGN\texttt{FITTER}. The cold gas component is shown in green, the stellar population component in yellow, the AGN accretion disc in blue, the AGN torus in purple (note that no torus component is needed in the model), and the total model in red. \textit{Top:} SED model obtained using the photometry of SDSS1335+0728 before the ZTF19acnskyy started generating alerts (prior to December 2019) from WISE, 2MASS, SDSS, and GALEX. This model (preSED-model) does not require an AGN torus component, but in most of the realisations it requires a faint AGN accretion disc component. \textit{Bottom:}  SED model obtained using the first \emph{Swift}/UVOT observation obtained in July 2021 (MJD 59426). For the infrared range we used the photometry from 2MASS and WISE obtained prior to December 2019, which allowed us to cover a larger range in wavelength. We also fixed the stellar age, SFH $\tau$, and $E(B-V)_{\text{gal}}$ parameters, using the results obtained in the preSED-model, letting the parameters of afterSED-model vary within the $1\sigma$ values obtained for preSED-model. This model (afterSED-model) does not require an AGN torus model, while it requires a much brighter AGN accretion disc component. }  \label{figure:sed_model}
\end{figure*}

The results of the SED fitting are shown in Figure \ref{figure:sed_model}. The top panel shows the results obtained for the preSED-model, while the bottom panel shows the results for the afterSED-model. From the figure, we note that none of the SED models require a torus component. Interestingly, we also note that most of the realisations of the preSED-model require a BBB component, although much fainter than the one required for the afterSED-model; however, we note that some realisations do not include this component, and instead, a younger stellar population is required. The obtained BBB components for the preSED-model and the afterSED-model are more dominant in the UV. According to the standard thin disc model \citep{Shakura73}, accretion discs around BHs are expected to have maximum effective temperatures proportional to $(R_{\text{Edd}}/M_{\text{BH}})^{0.25}$ (where $R_{\text{Edd}}$ is the Eddington rate of the source), and therefore for IMBHS and low-mass SMBHs, the accretion disc emission should peak in the far/extreme UV, and should contribute less flux in the optical range, compared to more massive BHs (see for instance Figure 18 in \citealt{Zou23}). This is consistent with what we observe in the modelled SEDs of SDSS1335+0728.

\begin{table*}[htpb]
  \begin{center}
    \caption{Output from AGN\texttt{FITTER} for SDSS1335+0728 before (preSED-model) and after (afterSED-model) ZTF19acnskyy started producing alerts. The reported values correspond to the median and $2\sigma$ errors.}
    \label{table:sed_out}
    \begin{tabular}{ccccccc} 
   
\hline
\hline

\rule{0pt}{3ex}  SED model &  SFH $\tau$ [Gyr] & age [log yr] & $E(B-V)_{\text{gal}}$ & SFR [$M_{\odot}$/yr] & $M_{\star}$ [log$_{10} M_{\odot}$] & $L_{\text{BBB}}$(0.05-1) [log$_{10}$ erg s$^{-1}$] \\[1ex]
\hline

\rule{0pt}{4ex}  preSED-model & $1.00_{-0.95}^{+0.96}$ & $10.02_{-0.32}^{+0.10}$ & $0.07_{-0.04}^{+0.07}$ & $0.34_{-0.13}^{+0.48}$ & $10.20_{-0.23}^{+0.14}$ & $42.21_{-4.37}^{+0.16}$ \\ [2ex]

\hline

\rule{0pt}{4ex}  afterSED-model & $0.85_{-0.52}^{+0.78}$ & $10.05_{-0.17}^{+0.05}$ & $0.10_{-0.02}^{+0.01}$ & $0.38_{-0.19}^{+0.57}$ & $10.15_{-0.14}^{+0.07}$ & $43.09_{-0.09}^{+0.05}$\\ [2ex]

\hline
\hline

  \end{tabular}
  \end{center}
\end{table*}

Table \ref{table:sed_out} shows the preSED-model and afterSED-model output parameters, including SFH $\tau$, age, $E(B-V)_{\text{gal}}$, SFR, the stellar mass ($M_{\star}$), and $L_{\text{BBB}}$(0.05-1). The table shows the median and the $2\sigma$ errors (computed using the 2.5 and 97.5 percentiles of the different SED realisations) for each parameter. The obtained stellar masses are in agreement with the value reported in the MPA-JHU catalogue. The median BBB luminosity in the preSED-model is $\sim1.62\times10^{42}$ erg s$^{-1}$s around one order of magnitude smaller than the one obtained for the afterSED-model ($\sim1.23\times10^{43}$ erg s$^{-1}$). However, we note that the measured BBB luminosity in the preSED-model for a younger stellar population is $7.84\times10^{37}$ erg s$^{-1}$ (which corresponds to the percentile 2.5 {of the preSED-model realisations}). If we assume that  $L_{\text{BBB}}$(0.05-1) is a good representation of the bolometric luminosity (since the upper limit of the X-ray emission is much fainter than the accretion disc luminosity), we can see that the bolometric luminosities measured from both SED models are consistent with previous observations of LLAGNs, including composite (or transitioning) galaxies, LINERs, and Seyfert 1s (e.g. \citealt{Ho09}).

Assuming the BH masses derived in Section \ref{section:selection}, we can estimate the Eddington rate $R_{\text{Edd}}$ of the source, using the definition presented in \cite{Netzer13}. By assuming a BH mass of $3.3\times10^{6} M_{\odot}$, prior to December 2019 we obtain $R_{\text{Edd}}$ of 0.003 ($1\times10^{-7}$ if we assume the lowest BBB luminosity), and after a value of 0.025, while if we assume a BH mass of $1.5\times10^{6} M_{\odot}$, we obtain $R_{\text{Edd}}$ of 0.007 ($3\times10^{-7}$ if we assume the lowest BBB luminosity) and 0.055 before and after December 2019. These $R_{\text{Edd}}$ estimated values are also consistent with previous observations of LLAGNs (e.g. \citealt{Ho09}). Moreover, the difference among these values is similar to what has been recently observed for the transition of CSAGNs \citep{Ruan19,Graham20,Guolo21,Temple23,LopezNavas23b}, and in the differences between LINERs and Seyfert 2s (e.g. \citealt{Hernandez-Garcia16}).

The potential detection of a BBB component prior to December 2019, and the SED and $R_{\text{Edd}}$ evolution of SDSS1335+0728, support the hypothesis of an SMBH that was accreting at very low accretion rates and that is currently turning on (AGN ignition event). However, the BBB luminosity measured in the afterSED-model is consistent with previous observations of TDEs \citep{Hammerstein23}, and thus the TDE scenario cannot be discarded.

\subsection{Spectroscopic analysis}\label{section:spec}

SDSS1335+0728 was observed prior to December 2019 by SDSS in MJD 54156. According to the SDSS pipeline classification, the source corresponds to a star-forming galaxy. From its emission line ratios, the source was classified as composite in the MPA-JHU catalogue and by \cite{Toba14}. Moreover, this spectrum has no coronal lines detected in the Coronal Line Activity Spectroscopic Survey (CLASS; \citealt{Reefe23}) catalogue. More recently, the source was observed by LAMOSTmin MJD 57155. The Reference Catalogue of Spectral Energy Distributions version 2 (RCSED2\footnote{\url{https://rcsed2.voxastro.org}; \citealt{Chilingarian12,Chilingarian17}}) classifies this spectrum as star-forming galaxy. From these observations, we can conclude that SDSS1335+0728 had either no nuclear activity or very low AGN activity before it started showing variations in late 2019. 

As mentioned in Section \ref{section:follow_data}, in mid-2021 we performed a spectroscopic follow-up of SDSS1335+0728, observing the source with SOAR/Goodman (MJD 59372 and 60337), VLT/X-shooter (MJD 59789), and Keck/LRIS (MJD 60137). 

\begin{figure*}[htbp!]
\begin{center}
\includegraphics[scale=0.55]{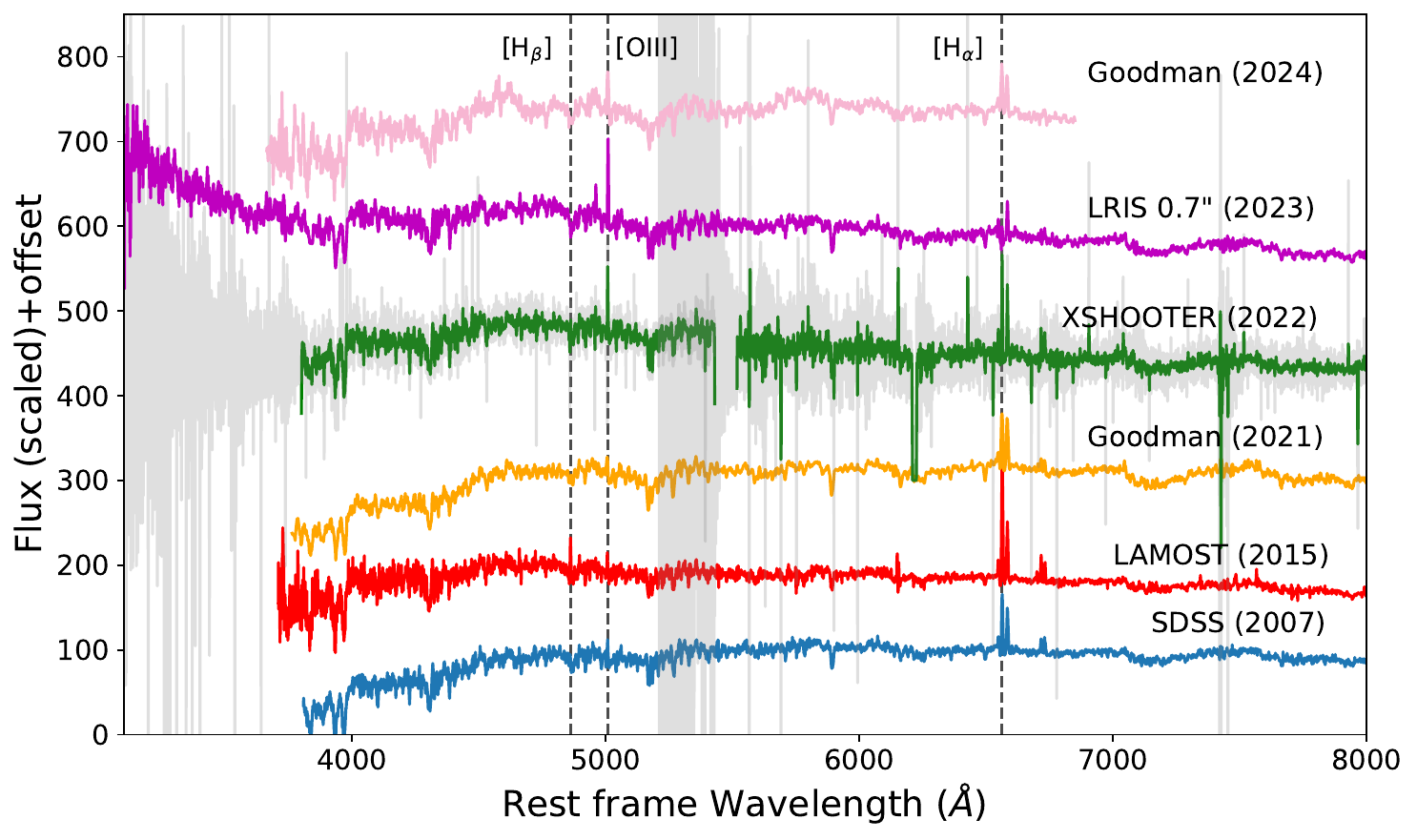}
\begin{tabular}{cc}
  \includegraphics[scale=0.5]{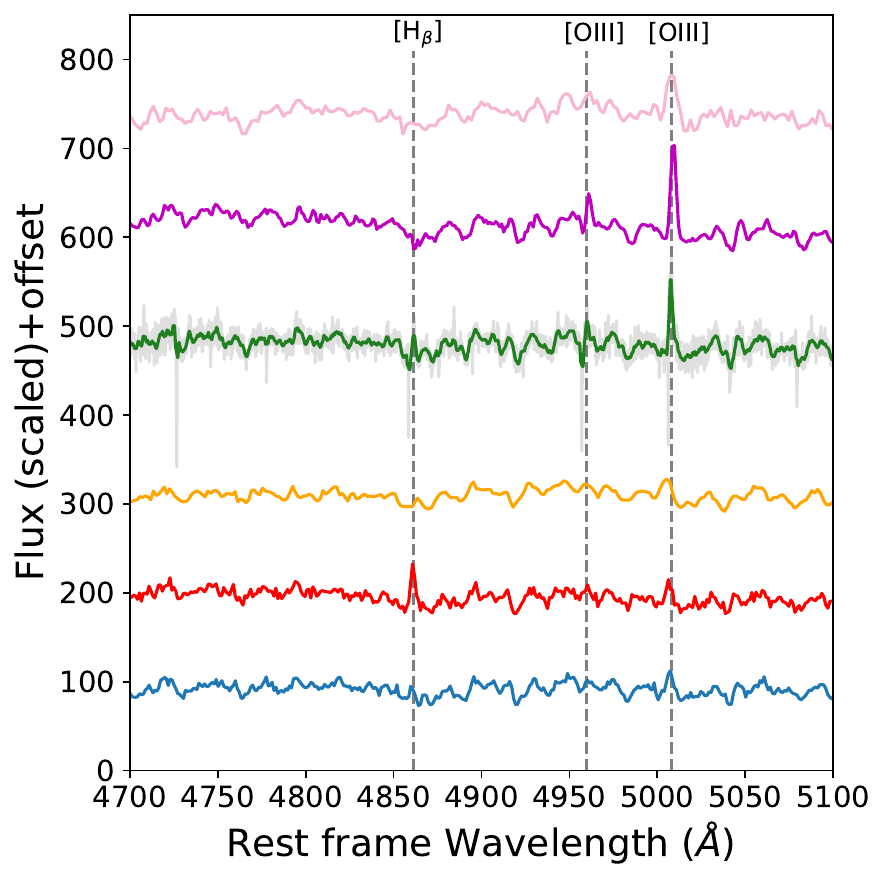}  &
  \includegraphics[scale=0.5]{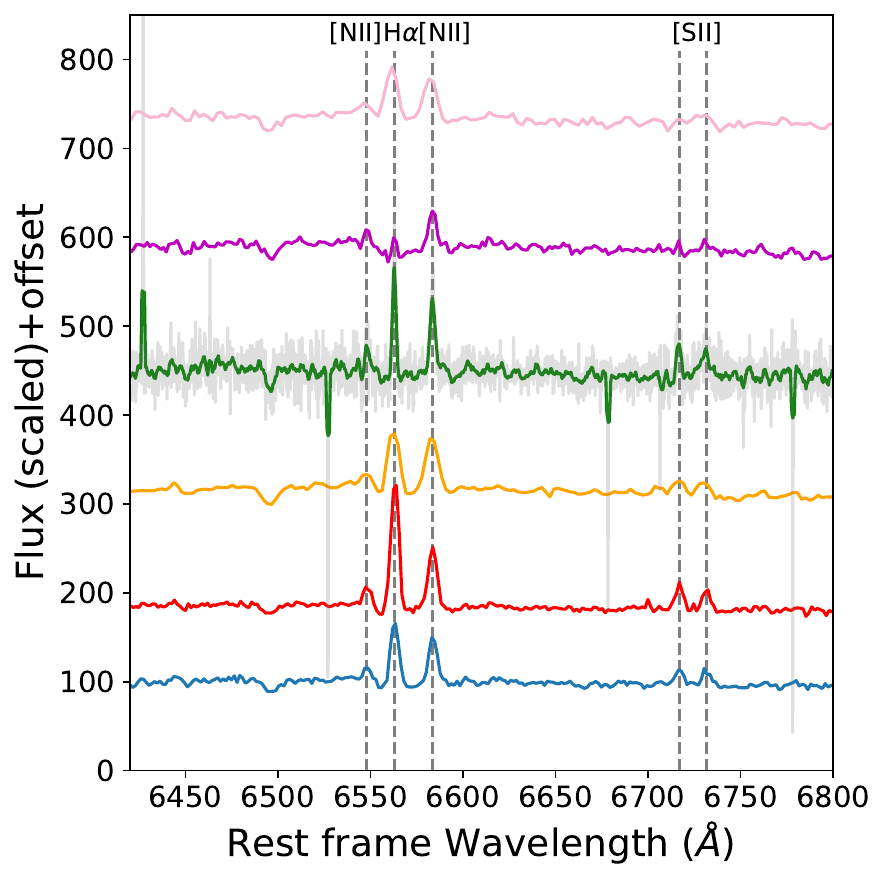} \\
\end{tabular}
\caption{Spectra of  SDSS1335+0728. \textit{Top:} Full spectra observed by SDSS (2007; blue), LAMOST (2015; red), Goodman (2021 and 2024; yellow and pink), X-shooter (2022; green), and LRIS (2023, 0\farcs7; purple). The original X-shooter spectrum is shown in grey and the box smooth (with a window of 10 pixels) binned spectrum is shown in green. \textit{Bottom:} The left panel shows a zoom around the H$_\beta$ line region, and the right panel shows a zoom around the H$_\alpha$ line region. \label{figure:spec_all}}
\end{center}
\end{figure*}

Six of the seven available spectra are shown in Figure\,\ref{figure:spec_all} (the LRIS 1\farcs5 slit spectrum is not presented, but is similar to the 0\farcs7 slit observation). The top panel shows the full spectra, while the bottom left and right panels show a zoom around the H$_{\beta}$ and H$_{\alpha}$ lines, respectively.

From a visual inspection of Figure \ref{figure:spec_all}, we note that most of the observations are dominated by the host-galaxy component of SDSS1335+0728 in the optical range, while the Keck LRIS spectrum shows a much more prominent UV emission (the same is observed in the 1\farcs5 slit spectrum), which is consistent with the observations obtained from \emph{Swift}/UVOT. The X-shooter and Goodman 2021 observations do not allow us to study wavelengths shorter than 3800 $\AA$ (because of the low signal-to-noise ratio at this wavelength range), and thus we cannot confirm spectroscopically if this UV emission was also present in 2021 and 2022; however, this is confirmed by the \emph{Swift}/UVOT observations. Moreover, we note that the relative strength of the [OIII] and [NII] forbidden lines with respect to the Balmer lines has changed with time. This change in the line fluxes could be due to changes in the fibre/slit sizes or to the evolution of the source (this is further discussed in Section \ref{section:spec_bpt}).

\subsubsection{Spectral modelling}\label{section:spec_model}
 
We modelled each spectrum using the Penalized PiXel-Fitting (pPXF) software (\citealt{2004PASP..116..138C}, \citealt{2017MNRAS.466..798C}). Each model includes stellar populations with a wide range of metallicity and age \citep{2016MNRAS.463.3409V}, emission lines (narrow and broad) and {a set of power laws of the form $(\lambda/\lambda_N)^{\alpha}$, where $\lambda$ is the wavelength, $\lambda_N=5000 \AA$ is a normalization factor and $\alpha$ the slope, that goes from -4 to 0 in steps of 0.1}. All the spectra, except that obtained by X-shooter, were fitted over the whole wavelength range that includes the H$_{\beta}$ and H$_{\alpha}$ regions. For the case of the X-shooter spectrum, to reduce the noise (see grey spectrum in the top panel of Figure \ref{figure:spec_all}), we used a box smooth\footnote{Using \texttt{specutils}: \url{https://specutils.readthedocs.io/en/stable/index.html.}} with a window of 10 pixels  (equivalent to $\sim$5 resolution elements at 7750\,\AA). This smoothed spectrum is shown in green in Figure \ref{figure:spec_all}. Then, we fitted the optical range that includes the H-$\alpha$ region and used these results to constrain the range of slopes for the set of power laws to fit the UV spectrum with the H-$\beta$ region. The range of slopes was limited to values from $-1.1$ to 0 and no other constraints were applied. The results of the pPXF modelling for all the spectra are shown in Appendix \ref{app:spec_model_plots}.

The best-fits from pPXF yield consistent ages of the stellar populations in all the spectra, which are also consistent with what we obtain from the SED modelling. The SDSS, LAMOST and SOAR 2021 spectra do not require an AGN power-law component. For the LRIS spectrum a power-law component with a slope of $-2.901$ ($-3.274$ for the 1\farcs5 slit) is needed. For the case of the X-shooter data, we modelled the spectrum with and without a power-law component, resulting in similar stellar population ages, and with similar residual sum of squares (RSS), with RSS of $4.15\times10^{-30}$ for the model with a power-law component and $4.13\times10^{-30}$ for the model without a power-law component. This prevented us from identifying the best model for the X-shooter spectrum. Thus, we decided to include the model with a power-law component in the analysis, as the \emph{Swift}/UVOT contemporaneous image shows a strong UV emission. For the SOAR 2021 spectrum, the model requires a power-law component with a slope of $-1.22$.  

We measured the uncertainties of the fits by performing Monte Carlo simulations. For this, we divided the best-fit residuals into five (six for the case of X-shooter) wavelength ranges limited by the H$_\beta$ and H$_\alpha$ regions. For each range, we compute the standard deviation and then use it to generate a set of random values from a Gaussian with the centre at zero, with these values we build a random simulated noise. Then, we added the simulated noise to the best-fit model and performed the fit with the same setting used to fit each spectrum. We used 1000 simulations and the 16 and 84 percentiles for the errors of measured values.

\subsubsection{Emission line diagnostics}\label{section:spec_bpt}

None of the available spectra of SDSS1335+0728 require a broad line emission component in the pPXF fits. Moreover, none of them require Bowen fluorescence (BF; \citealt{Netzer85}) emission lines, which have recently been found in flaring AGNs \citep{Trakhtenbrot19NatAs,Frederick21} and TDEs \citep{vanVelzen20,Zabludoff21}.   

\begin{figure*}[htbp!]
\begin{center}
\begin{tabular}{cc}
  \includegraphics[scale=0.48]{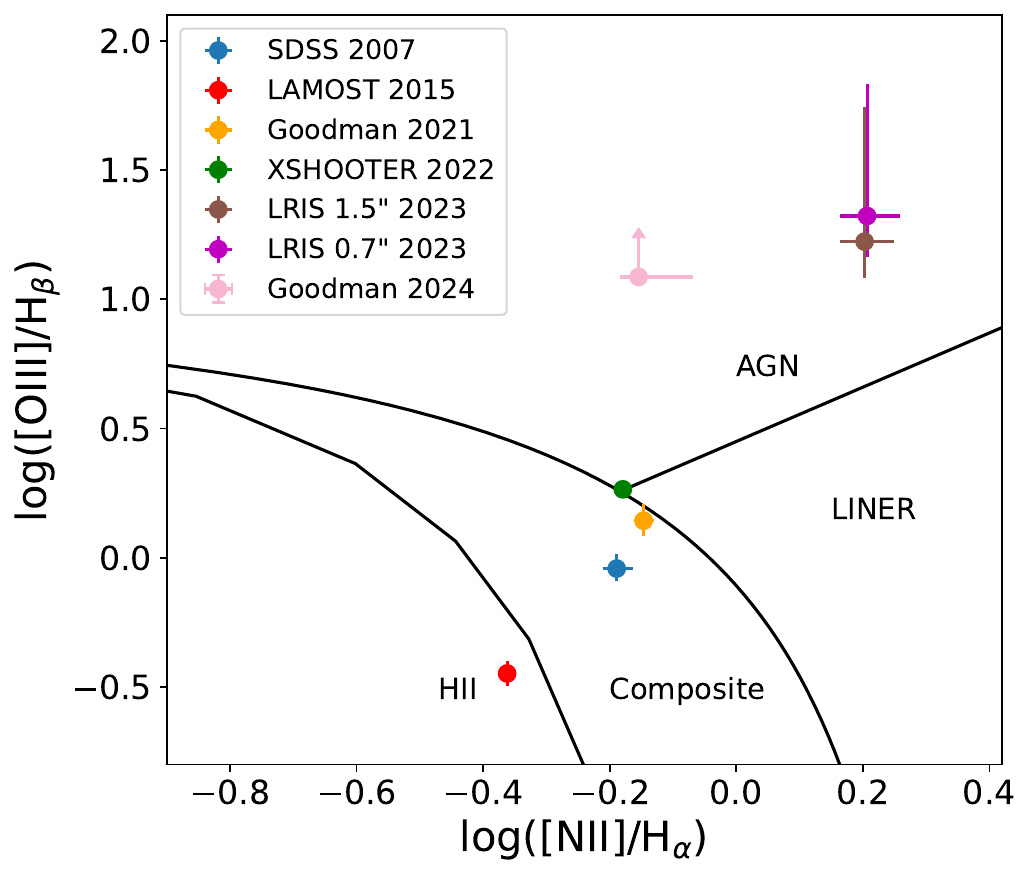}  &
  \includegraphics[scale=0.48]{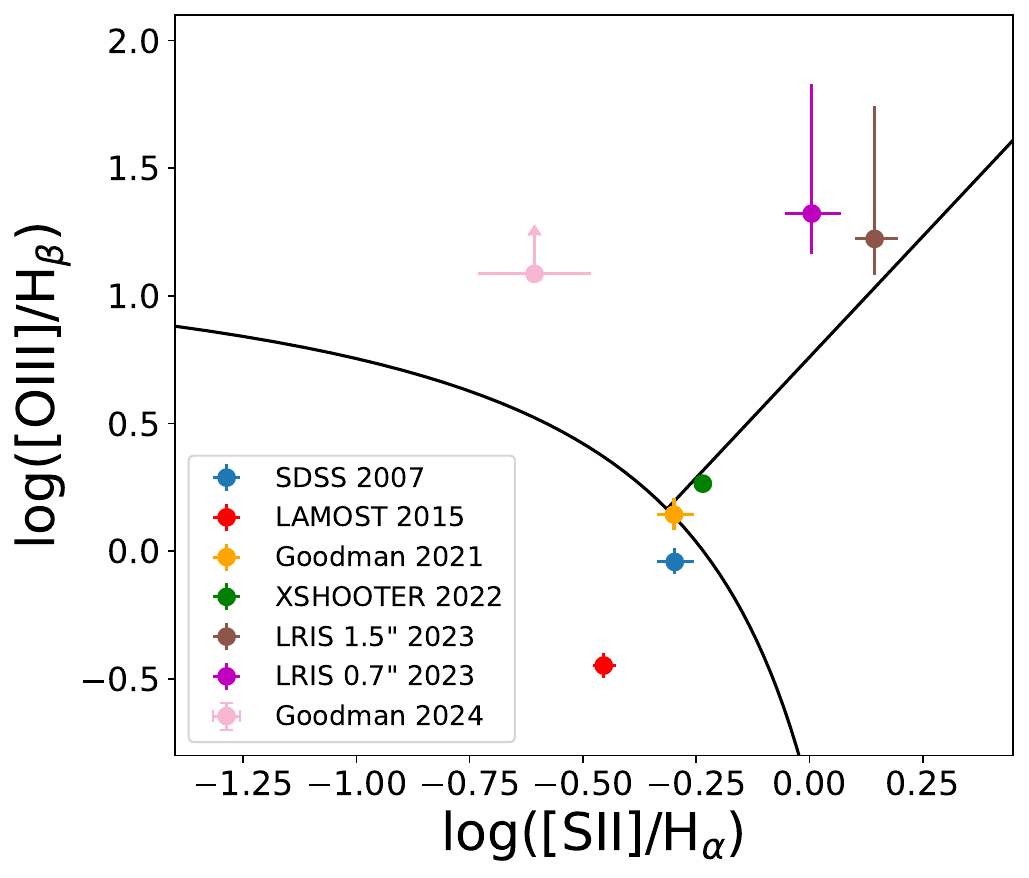} \\
\end{tabular}
\caption{BPT diagrams for all the modelled spectra. The left panel shows the [OIII]$\lambda$5007/H$_{\beta}$ versus [N II]$\lambda$6583/H$_{\alpha}$ diagnostic, while the right panel shows the [OIII]$\lambda$5007/H$_{\beta}$ versus  [SII]$\lambda$6716.47,6730.85/H$_{\alpha}$ diagnostic.  The solid black lines show the theoretical relations to separate the star-forming, composite and AGN (Seyfert and Liner) regions in both diagrams.   \label{figure:bpt_diagrams}}
\end{center}
\end{figure*}

To have a better idea of any possible AGN-like activity in SDSS1335+0728, we constructed BPT diagrams (after Baldwin, Phillips, and Terlevich; \citealt{Baldwin81}), consisting of indexes that relate the strengths of some narrow emission lines  (NELs; H$_{\beta}$, [O III]$\lambda$5007, H$_{\alpha}$, [N II]$\lambda$6583, and [S II]$\lambda$6716.47,6730.85) as proxies for the state of ionization of a region. From the best-fit {pPXF} results and the Monte Carlo simulations, we obtained the emission lines intensities (with their respective errors) to compute the ratios  [OIII]$\lambda$5007/H$_{\beta}$, [N II]$\lambda$6583/H$_{\alpha}$, and [SII]$\lambda$6716.47,6730.85/H$_{\alpha}$. These are shown in Figure\,\ref{figure:bpt_diagrams}.  We used the theoretical relations from \cite{2001ApJ...556..121K}, \cite{2003MNRAS.346.1055K}, \cite{2006MNRAS.372..961K} and \cite{2007MNRAS.382.1415S}, to separate star-forming, composite and AGN (Seyfert and LINER) regions of the two diagrams. 

For the SDSS and LAMOST spectra, we obtain results in agreement with RCSED2\footnote{\url{https://rcsed2.voxastro.org/data/obj/450785}}. The LAMOST spectrum is classified as star-forming, and the SDSS spectrum as composite. Differences in the fibre aperture could explain this difference in the classifications of the SDSS and LAMOST spectra, and worse seeing conditions normally observed at the location of LAMOST. For Goodman 2021 and X-shooter, the results show a preferred position near the limit between composite-AGN and star-forming-AGN, indicating a more energetic ionization state than the SDSS and LAMOST results. While for the case of the LRIS observations, a drastic evolution towards the AGN region of the BPT diagram is observed for both slits. The same is observed for the Goodman 2024 spectrum, but we see an evolution in the BPT location with respect to the LRIS observations. This suggests that the Narrow Line Region (NLR) of SDSS1335+0728 has had enough time to react to the increased ionising continuum after more than 3.6\,years of activity (the time between the first ZTF alert and the LRIS spectra).

\begin{figure}[htbp!]
\begin{center}

\includegraphics[scale=0.6]{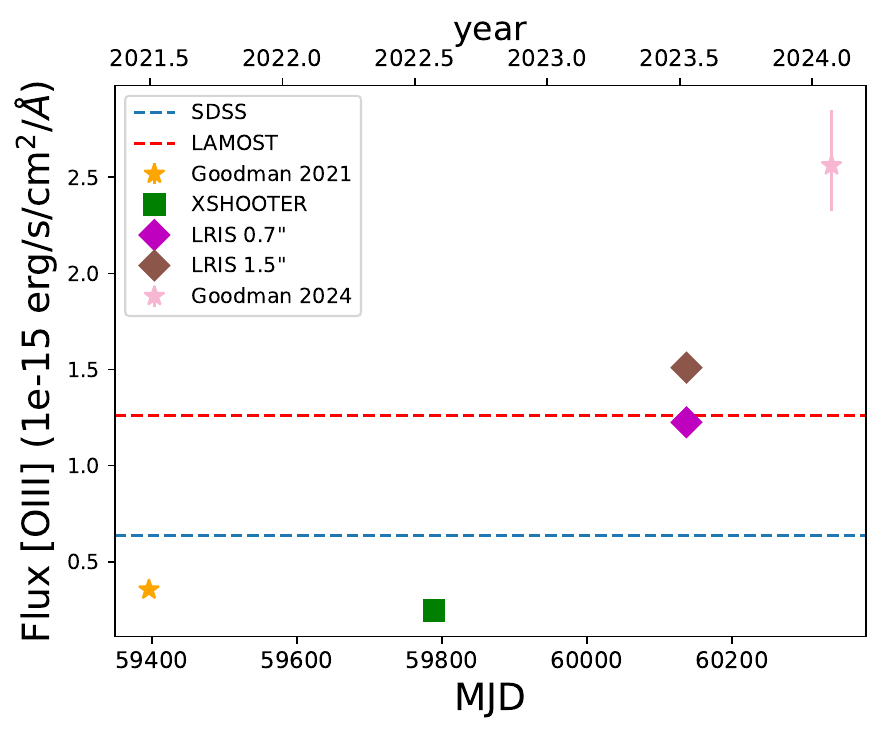}  

\caption{Light curve of the [OIII] NEL of SDSS1335+0728 for observations obtained after December 2019. We include the measurements obtained from Goodman (yellow and pink stars), X-shooter (green square), and LRIS (magenta and brown diamonds). The horizontal dashed lines show the flux level of the SDSS and LAMOST observations. \label{figure:line_lc}}

\end{center}
\end{figure}

Figure\,\ref{figure:line_lc} shows the evolution of the  [OIII] NEL after December 2019, which is the line that shows the most evident evolution among the different epochs. We include as a reference the flux level of the line in the SDSS and LAMOST spectra. These fluxes are highly dominated by the host-galaxy as they were obtained using $3''$ and $3.''5$ fibres, respectively. All the observations were obtained with different instruments, slit apertures, and different seeing qualities (see Table \ref{table:spec_sum}). Thus, the observed evolution should be considered with caution. We see a drastic increment in the [OIII] flux in both LRIS observations (of $\sim$ two to four times the flux observed by Goodman 2021 and X-shooter) and in the Goodman 2024 spectrum. The level of the LRIS and Goodman 2024 [OIII] flux is also larger than the ones obtained from SDSS and LAMOST prior to 2019. Thus, this increase in the [OIII] flux is difficult to explain by variations in the slit size or by the use of a different instrument.

The evolution of the [OIII] line can be explained by the delayed reaction of the NLR to the increase in the ionising flux since December 2019. From the BPT diagram evolution, and the [OIII] line flux variability, we can estimate an upper limit for the innermost radius of the NLR of 3.6 light years (or 1.1 pc). This is consistent with the results of \cite{Peterson13}, who measured a radius of 1--3\,pc for NGC 5548 (BH mass of $7\times10^{7} M_{\odot}$; bolometric luminosity of $2.82\times10^{44}$ erg s$^{-1}$ \citealt{Ebrero16}). Generally, the NLR size is thought to span hundreds to thousands of parsecs \citep{Netzer13}. However, when a source is just activating, the inner parsec of ionised gas should react first. This would be possible if there is an extensive zone in the NLR of relatively low filling factor located close to the nucleus, from which most of the nuclear OIII emission of SDSS1335+0728 is coming from. Such inhomogeneous density in the NLR has been previously observed in other objects, like NGC1068 \citep{Kraemer98}.

The spectral analysis supports the hypothesis of an AGN that is turning on, whose NLR is starting to react to the X-ray/UV/optical emission. However, significant NEL variability has also been detected in TDEs (e.g. \citealt{Charalampopoulos22}). The lack of BELs, along with a strong rise in the UV, could be explained if there are few BLR clouds able to produce strong enough broad-emission lines \citep{Osterbrock06}. {As mentioned in Section \ref{section:mir_var}, one of the possible scenarios to explain the MIR flux evolution is that a torus component is being formed.} This implies that clouds may be falling in, populating the torus first and eventually the BLR; thus, we might detect BELs in the future. Further monitoring will be necessary to confirm this hypothesis.

\section{Conclusions}\label{section:conclusion}

SDSS1335+0728 started triggering alerts in the ZTF alert stream in December 2019 (object ID ZTF19acnskyy), and was classified as an AGN by the ALeRCE LCC. The source did not show variations in the UV/optical/infrared range for almost two decades before this date. From the behaviour of SDSS1335+0728, we considered two hypotheses for its origin: (a) a $\sim 10^6 M_{\odot}$ BH that has now turned on as an AGN; and (b) a TDE (or another class of still unknown nuclear transient) in a $\sim 10^6 M_{\odot}$ BH. In 2021, we started a photometric and spectroscopic follow-up campaign to better understand the origin of the observed variations.

The classical TDE scenario is less favoured to explain the origin of the nuclear variation in SDSS1335+0728 based on the following findings: (a) The light curve of ZTF19acnskyy does not follow the smooth power-law decay typically observed in TDEs; (b) ZTF19acnskyy  had an absolute peak in the $g$ band of -16.9, and has been varying stochastically for more than 1550 days, contrary to what is expected from faint TDEs; (c) the SED shows a bluer-when-brighter behaviour, while TDEs are expected to show no colour variations; (d) there is no evidence of broad emission lines; and (e) the [OIII] line flux is quite prominent, and has increased with time, contrary to other TDEs observed in the optical range. 

Despite this, we cannot rule out the TDE hypothesis, as the variations observed in the nucleus of SDSS1335+0728 could have originated from an exotic TDE. In particular, some TDEs have shown a lack of broad emission lines, slow temporal evolution, and faint optical emission, such as eRASSt J074426.3+291606 \citep{Malyali23}. Others have shown optical, UV, or X-ray rebrightening (e.g. \citealt{Soraisam22,Malyali23a,Wevers23,Liu24}). Delayed X-ray emission is also common in TDEs, although not with the same timescales observed in SDSS1335+0728 \citep{Guolo24,Wevers24}. Moreover, we still do not clearly understand the behaviour of TDEs in already active AGNs \citep{Chan19}; however, in the last few years, some of these events have been discovered. For instance, \cite{Merloni15} proposed that the CSAGN discovered by \cite{LaMassa15} could be explained by a TDE in an accretion disc. Later, \cite{Trakhtenbrot19} and \cite{Ricci20} presented the discovery of 1ES 1927+654, a CSAGN, and claimed that a TDE in a type 2 AGN provoked an increase in the accretion rate at the innermost regions of the accretion disc. More recently, \cite{Homan23} presented the discovery of eRASSt J234402.9-352640 using eROSITA observations and concluded that a TDE within a turned-off AGN best matches their observations. All these examples, however, show temporal evolution similar to that observed in other TDEs.

We propose that a plausible origin for the nuclear activity observed in SDSS1335+0728 would be a $\sim 10^6 M_{\odot}$ BH that is now turning on as an AGN. This is supported by: (a) its stochastic variable emission; (b) the bluer-when-brighter behaviour; (c) the SED modelling, which is consistent with a BBB component that increased its flux by at least one order of magnitude, with a BBB luminosity of $\sim10^{43}$ erg s$^{-1}$ after December 2019, consistent with an AGN; (d) the estimated $R_{\text{Edd}}$, which has varied from less than $10^{-3}$ to $\sim10^{-2}$, consistent with previously observed CSAGNs; (e) the BPT diagram evolution from star-forming to AGN-like emission line ratios; (f) the [OIII] line evolution, which is responding to the UV/optical variations, and suggests that the NLR is starting to react to the increased ionising flux $\sim3.6$ years after the first ZTF alert, which implies a compact NLR; (g) a delayed MIR flux increase from about June 2022, and the redder W1-W2 colour, which could imply that a dusty torus is now being formed; and (h) the recent X-ray detection, 1514 days after the first ZTF alert, which shows very soft emission, similar to that observed in 1ES 1927+654 when its X-ray corona began to reform \citep{Ricci20}, which could imply that an X-ray corona is establishing itself in the nucleus of SDSS1335+0728. 

The lack of BELs suggests that SDSS1335+0728 does not yet have a BLR, as we do see a strong new UV continuum and relatively rapid optical variability, and therefore cannot associate the lack of BELs with obscuration effects. The recent MIR evolution cannot be explained by a preexisting {(AGN-like)} dusty torus, which suggests that a torus could be under formation, implying that clouds are falling in from the outside, populating the torus first and then the BLR. If this hypothesis is correct, we may eventually see BELs. If the source forms these structures in the future, it would be a perfect laboratory with which to study and test the different physical mechanisms proposed to explain the origin and evolution of the BLR and the torus in AGNs (e.g. \citealt{Czerny11,Elitzur14,Czerny17}). 

Unfortunately, disentangling the origin of the variability detected in the nucleus of SDSS1335+0728 is not yet possible. Future observations will allow us to understand whether SDSS1335+0728 will eventually fade or will continue evolving to become an AGN. In particular, multi-wavelength monitoring of the source is needed to see if a typical corona, BELs, and a dusty torus are eventually established, and to rule out a potential TDE or nuclear transient. X-ray monitoring and millimetre observations will be crucial to understand whether or not the recently detected X-ray emission originates from an X-ray corona under formation \citep{Ricci20,Ricci23b}. If this is the case, we may expect to detect hard X-ray emission in the future. High-angular-resolution integral field unit (IFU) spectroscopy would be ideal for studying the evolution of AGN signatures in spatially resolved emission-line diagnostic diagrams (e.g. \citealt{Mezcua20}) and for understanding the connection between the nuclear variations and the host properties. Moreover, observations in the $\sim800$ to $\sim3500$ $\AA$ UV range would help us to better understand the evolution of the ionising continuum and its connection with the evolution of the NELs. Finally, photometric and spectroscopic observations in the optical and infrared regimes would be key to studying the potential formation of BELs or a dusty torus, respectively.

\begin{acknowledgements}

The authors acknowledge support from the National Agency for Research and Development (ANID) grants: Millennium Science Initiative Program ICN12\_12009 (PSS,LHG,FEB,FF,MPH,AMMA), Millenium Nucleus NCN19\_058 (SB,PA,PL,MLMA), CATA-BASAL - FB210003 (FEB,CR,RJA), BASAL FB210005 (AMMA), FONDECYT Regular - 1241005 (FEB), 1230345 (CR), 1201748 (PL), 1200710 (FF), 1231718 (RJA). SB acknowledges financial support from ANID Becas 21212344, and the hospitality of ESO under the SSDF Project 28/23D. AB acknowledges financial support from Deutsche Forschungsgemeinschaft (DFG, German Research Foundation) under Germany's Excellence Strategy – EXC 2094 – 39078331. DH acknowledges support from DLR grant FKZ 50 OR 2003. MK acknowledges support from DLR grant FKZ 50 OR 2307. DT acknowledges support by DLR grant FKZ 50 OR 2203. AM acknowledges support by DLR under the grant 50 QR 2110 (XMM\_NuTra). PA and SB acknowledge support from the Max-Planck Society through a Partner Group grant. MJA acknowledges support from the NSF Grant \#2108402.

This work has been possible thanks to the use of AWS-U.Chile-NLHPC credits.

Powered@NLHPC: This research was partially supported by the supercomputing
infrastructure of the NLHPC (ECM-02).

Based on observations obtained with the Samuel Oschin 48-inch Telescope at the Palomar Observatory as part of the \textit{Zwicky} Transient Facility project. ZTF is supported by the National Science Foundation under Grant No. AST-1440341 and a collaboration including Caltech, IPAC, the Weizmann Institute for Science, the Oskar Klein Center at Stockholm University, the University of Maryland, the University of Washington, Deutsches Elektronen-Synchrotron and Humboldt University, Los Alamos National Laboratories, the TANGO Consortium of Taiwan, the University of Wisconsin at Milwaukee, and Lawrence Berkeley National Laboratories. Operations are conducted by COO, IPAC, and UW.

Based on observations obtained with the Samuel Oschin Telescope 48-inch and the 60-inch Telescope at the Palomar Observatory as part of the \textit{Zwicky} Transient Facility project. ZTF is supported by the National Science Foundation under Grants No. AST-1440341 and AST-2034437 and a collaboration including current partners Caltech, IPAC, the Weizmann Institute for Science, the Oskar Klein Center at Stockholm University, the University of Maryland, Deutsches Elektronen-Synchrotron and Humboldt University, the TANGO Consortium of Taiwan, the University of Wisconsin at Milwaukee, Trinity College Dublin, Lawrence Livermore National Laboratories, IN2P3, University of Warwick, Ruhr University Bochum, Northwestern University and former partners the University of Washington, Los Alamos National Laboratories, and Lawrence Berkeley National Laboratories. Operations are conducted by COO, IPAC, and UW.

We acknowledge the use of public data from the \emph{Swift} data archive (Obs IDs 00014350003, 00014350004, 00014350005, 00014350007).

The scientific results reported in this article are based in part on observations made by the \emph{Chandra} X-ray Observatory, observation ID:   29355.

Based on observations collected at the European Southern Observatory under ESO programme 109.24F5.001.

Based on observations obtained at the Southern Astrophysical Research (SOAR) telescope, which is a joint project of the Minist\'{e}rio da Ci\^{e}ncia, Tecnologia e Inova\c{c}\~{o}es (MCTI/LNA) do Brasil, the US National Science Foundation’s NOIRLab, the University of North Carolina at Chapel Hill (UNC), and Michigan State University (MSU), through CNTAC proposal CN2021A-17.

Some of he data presented herein were obtained at the W. M. Keck Observatory, which is operated as a scientific partnership among the California Institute of Technology, the University of California and the National Aeronautics and Space Administration. The Observatory was made possible by the generous financial support of the W. M. Keck Foundation.

The authors wish to recognize and acknowledge the very significant cultural role and reverence that the summit of Maunakea has always had within the indigenous Hawaiian community.  We are most fortunate to have the opportunity to conduct observations from this mountain.

This research has made use of the NASA/IPAC Extragalactic Database (NED), which is funded by the National Aeronautics and Space Administration and operated by the California Institute of Technology.

Funding for SDSS-III has been provided by the Alfred P. Sloan Foundation, the Participating Institutions, the National Science Foundation, and the U.S. Department of Energy Office of Science. The SDSS-III web site is http://www.sdss3.org/.

SDSS-III is managed by the Astrophysical Research Consortium for the Participating Institutions of the SDSS-III Collaboration including the University of Arizona, the Brazilian Participation Group, Brookhaven National Laboratory, Carnegie Mellon University, University of Florida, the French Participation Group, the German Participation Group, Harvard University, the Instituto de Astrofisica de Canarias, the Michigan State/Notre Dame/JINA Participation Group, Johns Hopkins University, Lawrence Berkeley National Laboratory, Max Planck Institute for Astrophysics, Max Planck Institute for Extraterrestrial Physics, New Mexico State University, New York University, Ohio State University, Pennsylvania State University, University of Portsmouth, Princeton University, the Spanish Participation Group, University of Tokyo, University of Utah, Vanderbilt University, University of Virginia, University of Washington, and Yale University.

\end{acknowledgements}

\bibliographystyle{aa}
\bibliography{bibliography.bib}

\begin{appendix}
\onecolumn
\section{Spectral modelling}\label{app:spec_model_plots}

The following figures (from \ref{figure:spec_sdss} to \ref{figure:spec_goodman24}) show the results of the pPXF spectral modelling. Each figure shows in the top panel the original spectrum in blue, the best fit (including all the components) in orange, the stellar component in green, the NELs in red, and the power law component in purple. The bottom panel of each figure shows the residual of the modelling (the difference between the original spectrum and the best fit).

\FloatBarrier

\begin{figure*}[hptb!]
\begin{center}
\begin{tabular}{cc}
  \includegraphics[width=0.51\linewidth]{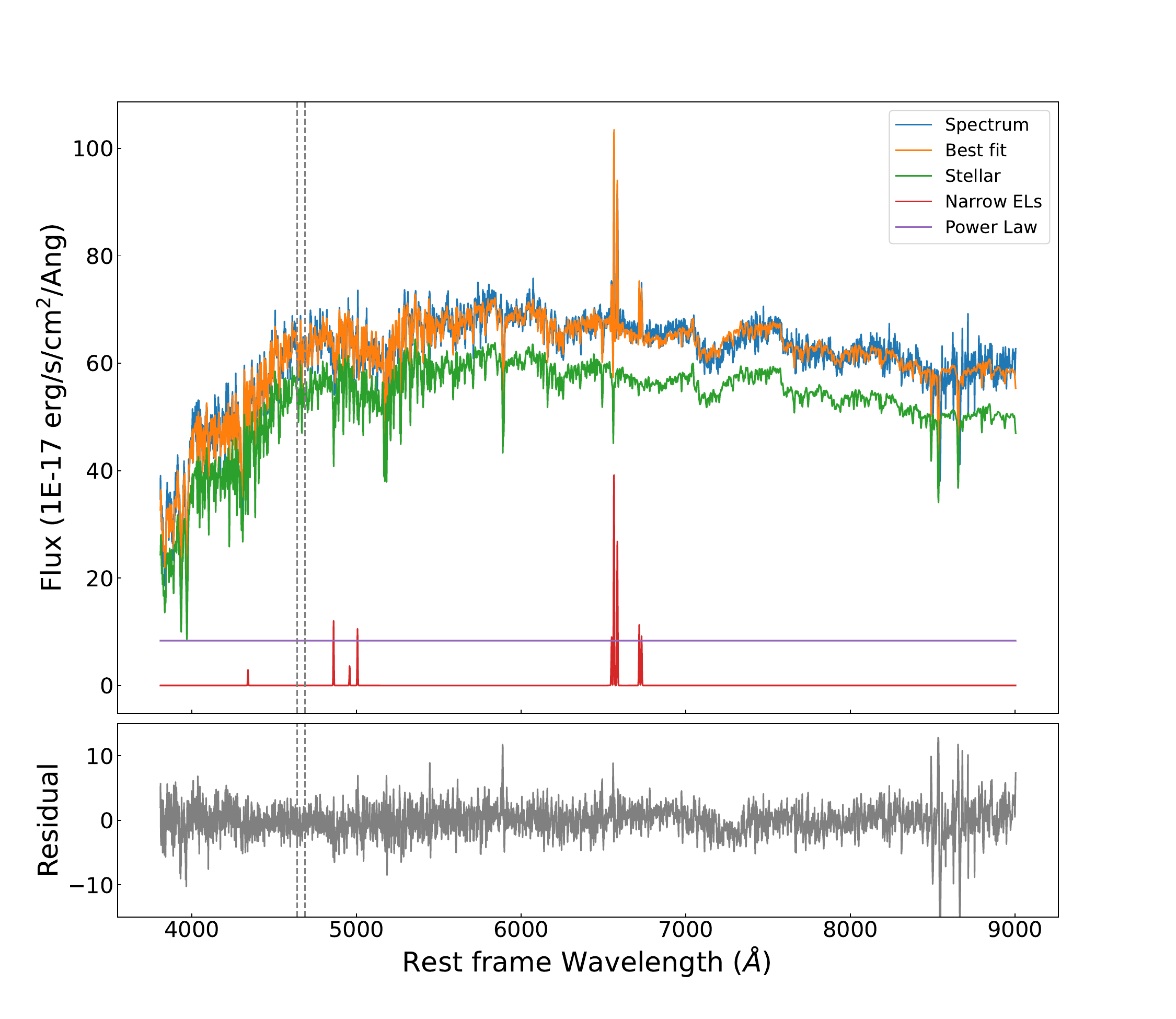} &
 \includegraphics[width=0.51\linewidth]{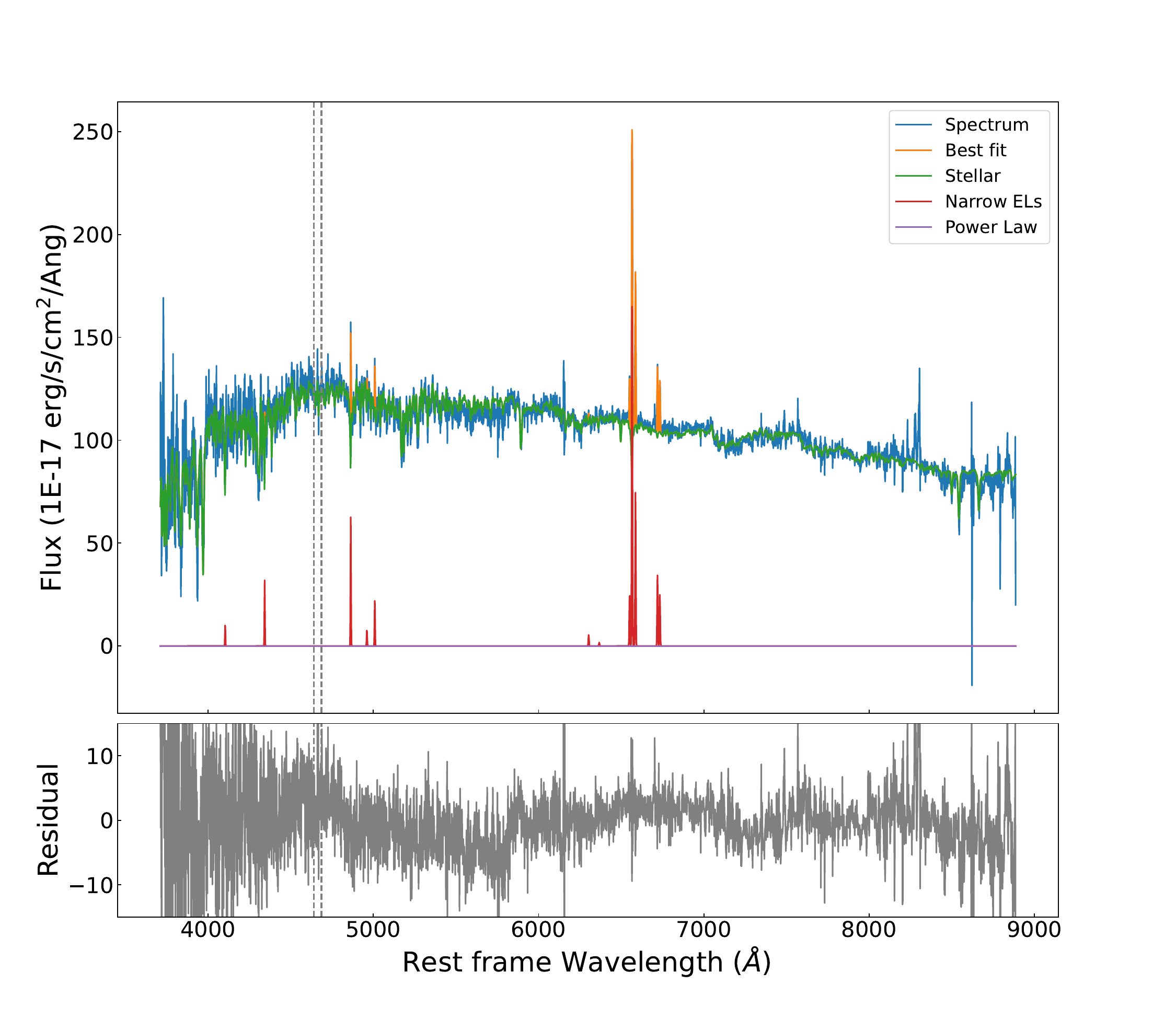} \\
\end{tabular}
\caption{Spectral modelling of the 2007 SDSS spectrum (MJD 54156; left) and of the 2015 LAMOST spectrum (MJD 57155; right). \textit{Top:} The original spectrum is shown in blue, the best fit (including all the components) in orange, the stellar component in green, the NELs in red, and the power law component in purple. \textit{Bottom:} The modelling residual is shown in grey. In both panels, the grey dashed lines show the location of the N III $\lambda4640$ and the He II $\lambda4686$ emission lines.
\label{figure:spec_sdss}}
\end{center}
\end{figure*}

\begin{figure*}[]
\begin{center}
\begin{tabular}{cc}
   \includegraphics[width=0.51\linewidth]{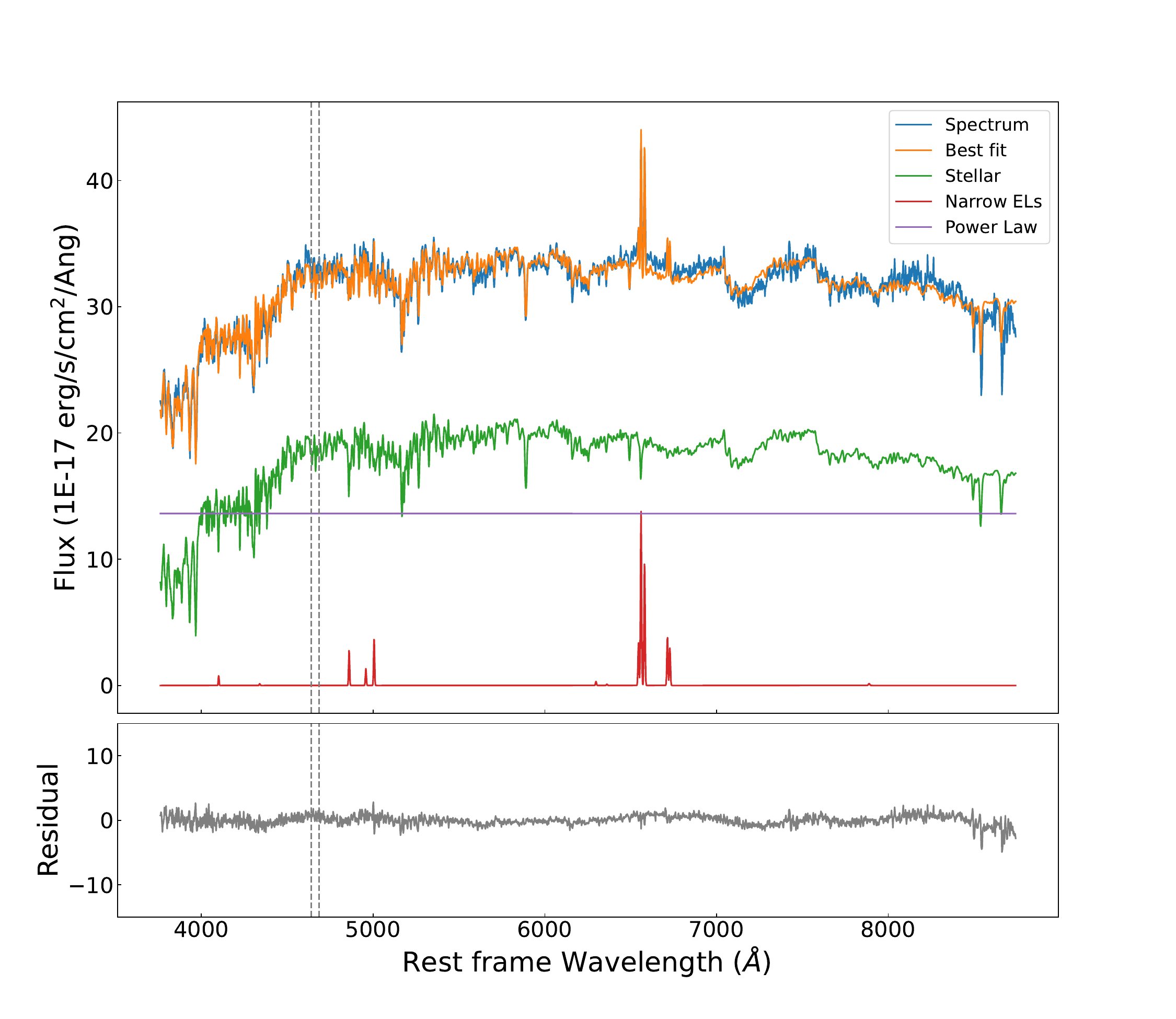} & \includegraphics[width=0.51\linewidth]{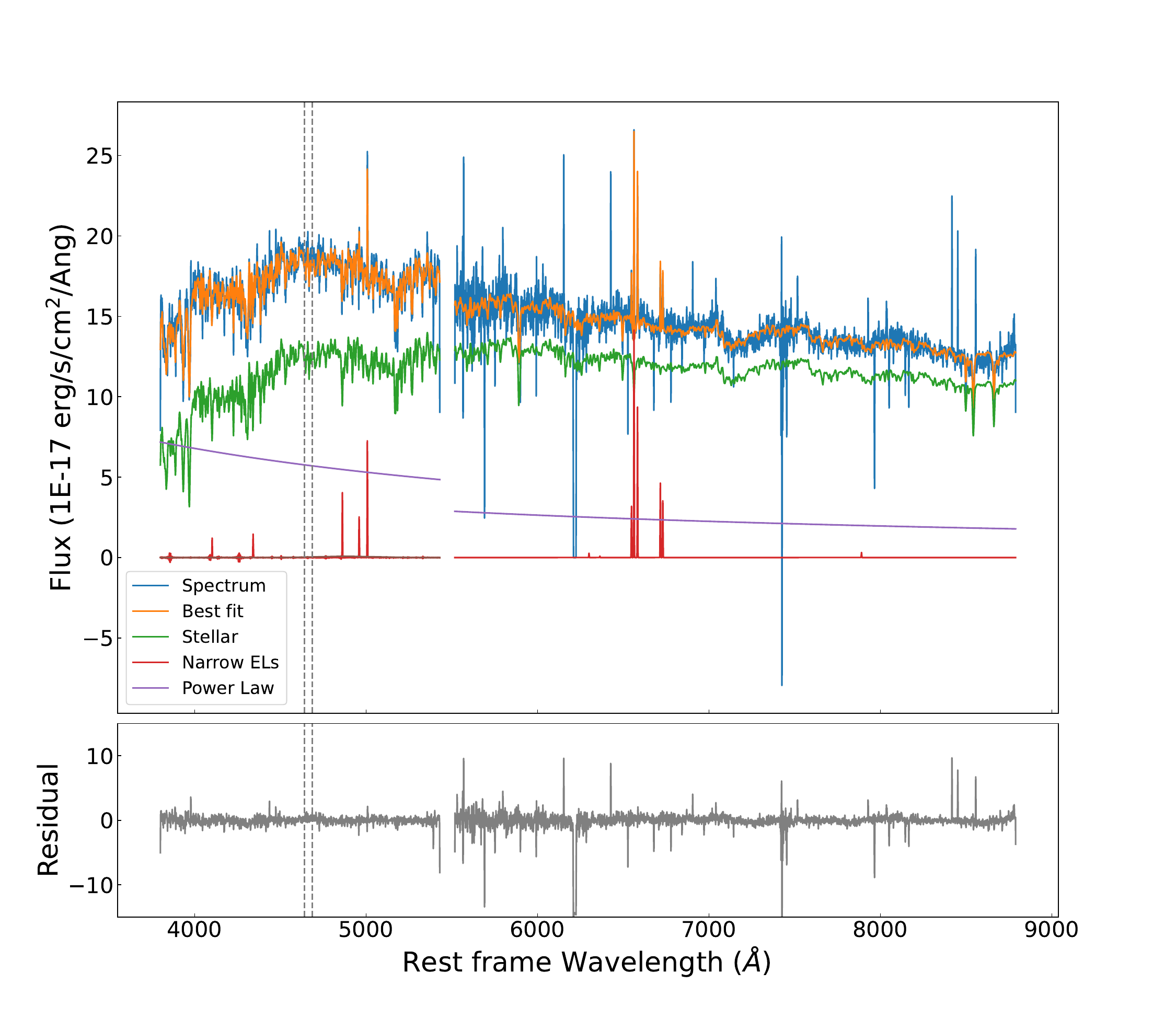} \\   
\end{tabular}
\caption{Same as Figure \ref{figure:spec_sdss} but for the 2021 Goodman spectrum (MJD 59396; left) and the 2022 X-shooter spectrum (MJD 59789; right).
\label{figure:spec_goodman21}}
\end{center}
\end{figure*}

\begin{figure*}[]
\begin{center}
\begin{tabular}{cc}
   \includegraphics[width=0.51\linewidth]{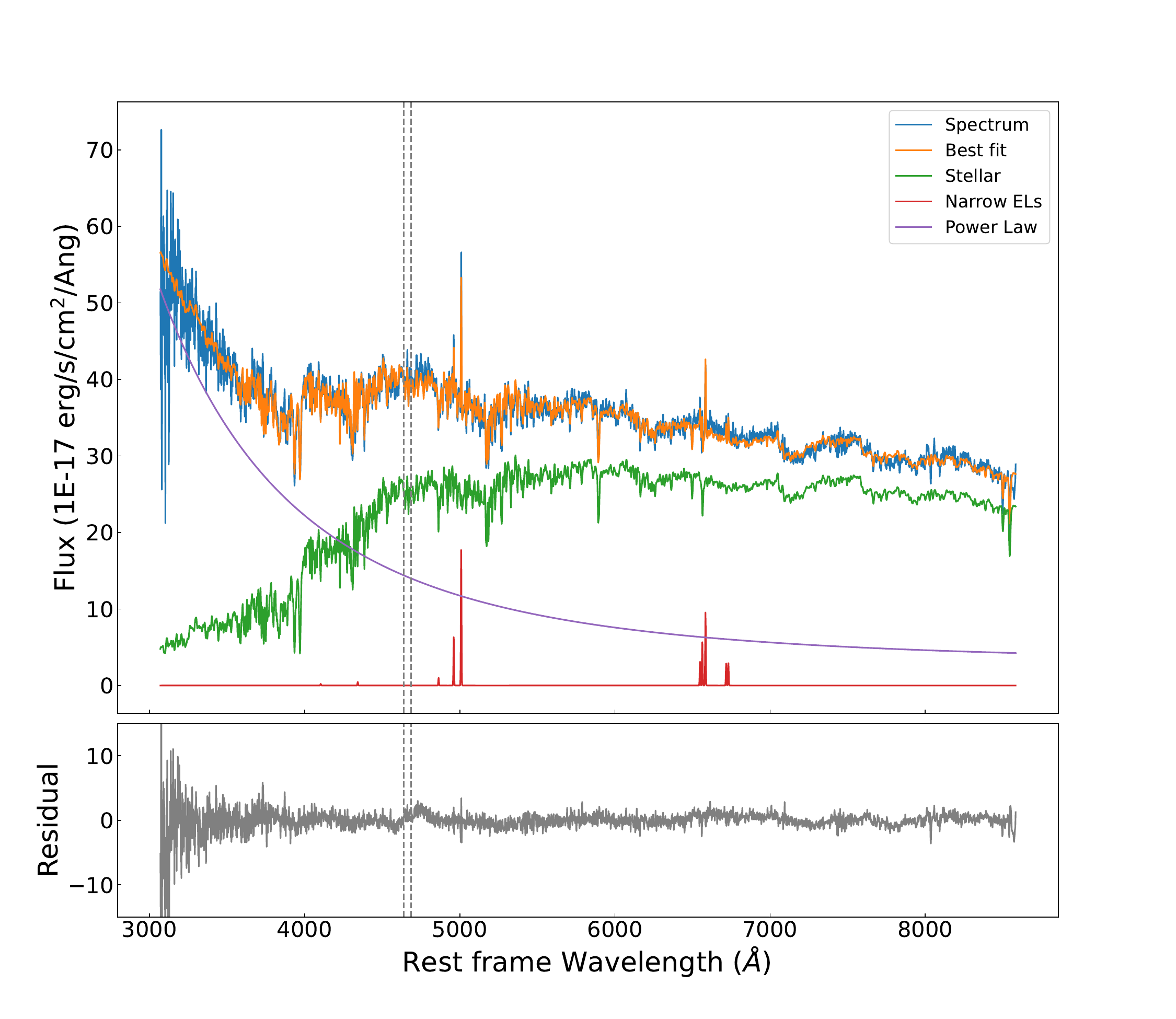} &
   \includegraphics[width=0.51\linewidth]{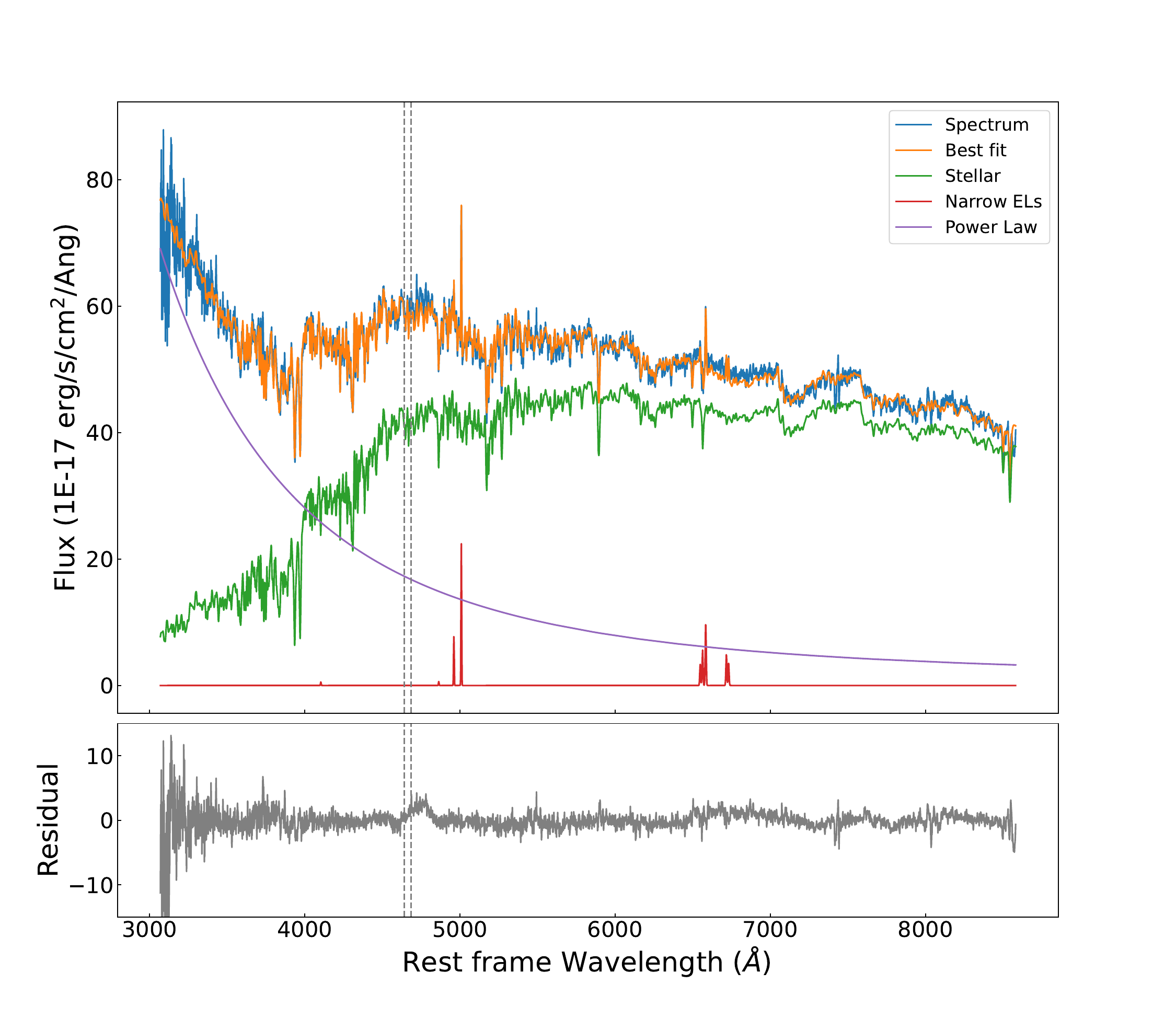} \\
\end{tabular}
\caption{Same as Figure \ref{figure:spec_sdss} but for the 2023 LRIS (MJD 60137) 0$''$.7 slit spectrum (left) and the 1$''$.5 slit spectrum.
\label{figure:spec_keck07}}
\end{center}
\end{figure*}

\begin{figure*}[]
\begin{center}
   \includegraphics[width=0.51\linewidth]{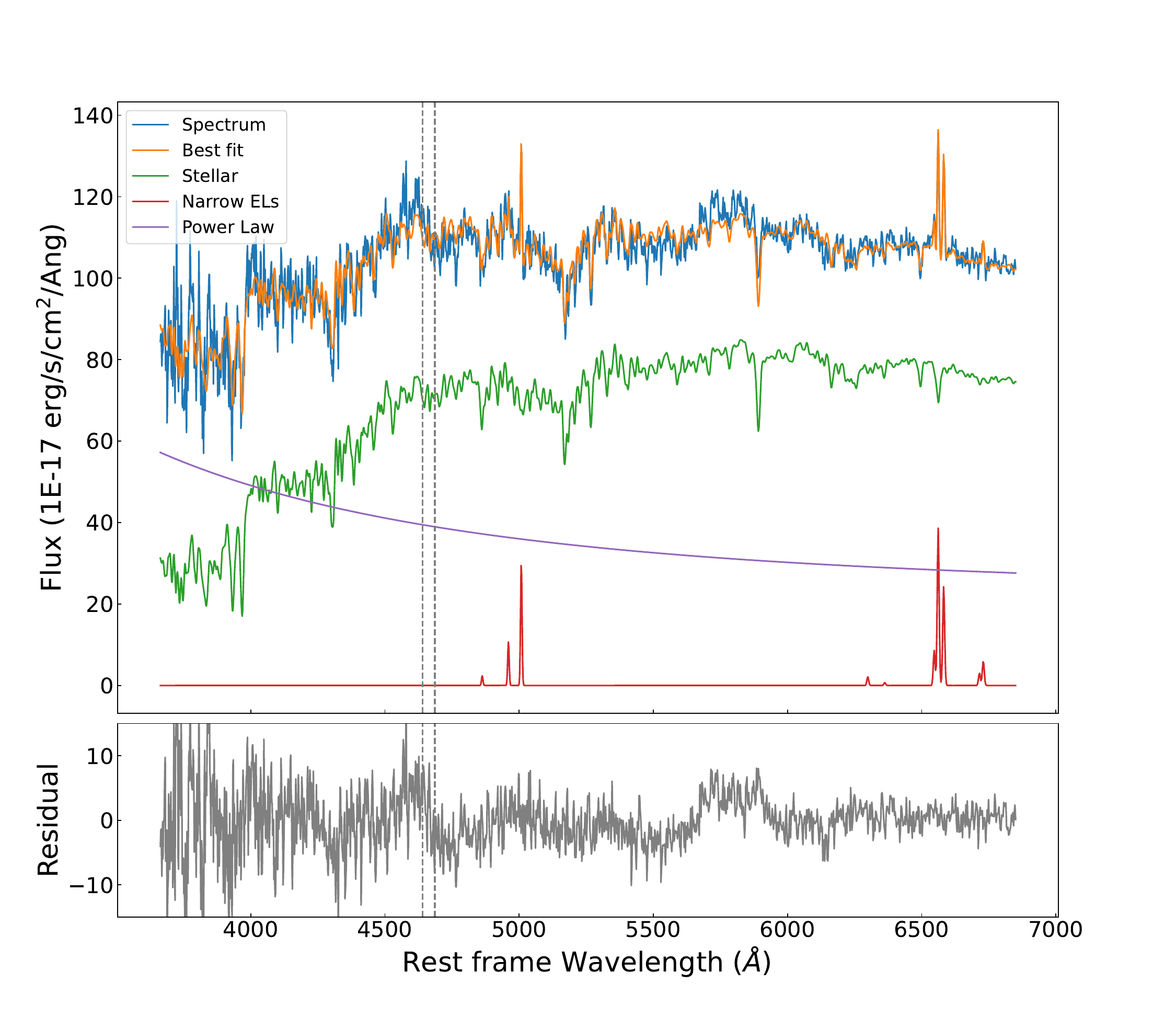} 

\caption{Same as Figure \ref{figure:spec_sdss} but for the 2024 Goodman spectrum (MJD 60337).
\label{figure:spec_goodman24}}
\end{center}
\end{figure*}

\end{appendix}

\end{document}